\documentclass[11pt]{article}
\usepackage{amsmath,amssymb,epsfig,bm,mathrsfs}
\usepackage{feynmp,slashed,color,mathrsfs,nicefrac,exscale}
\usepackage[colorlinks=true,linkcolor=blue,citecolor=blue,urlcolor=blue]{hyperref}
\newcommand{\bea}{\begin{eqnarray}}
\newcommand{\eea}{\end{eqnarray}}
\newcommand{\be}{\begin{equation}}
\newcommand{\ee}{\end{equation}}
\newcommand{\Omegavec}{{\bm \Omega}}
\newcommand{\vecsigma}{{\bm \sigma}}
\newcommand{\vecnabla}{{\bm \nabla}}
\newcommand{\vecnu}{{\bm \nu}}
\newcommand{\vectau}{{\bm \tau}}

\newcommand{\vecp}{{\bm p}}
\newcommand{\vecq}{{\bm q}}
\newcommand{\vecf}{{\bm f}}
\newcommand{\vecl}{{\bm l}}
\newcommand{\vecv}{{\bm v}}
\newcommand{\vecJ}{{\bm J}}
\newcommand{\vecr}{{\bm r}}

\newcommand{\vecA}{{\bm A}}
\newcommand{\vecM}{{\bm M}}

\newcommand{\vecgamma}{{\bm\gamma}}
\newcommand{\ep}{\varepsilon}

\newcommand {\pp} {\parallel}
\newcommand {\pin} {\rm pin}
\newcommand {\mf} {\rm mf}
\newcommand {\mg} {\rm mag}

\newcommand{\beq}{\begin{eqnarray}}
\newcommand{\eeq}{\end{eqnarray}}
\newcommand{\n}{\mathrm{n}}
\newcommand{\e}{\mathrm{e}}
\newcommand{\p}{\mathrm{p}}
\newcommand{\x}{\mathrm{x}}
\newcommand{\y}{\mathrm{y}}
\newcommand{\vv}{\mathrm{v}}
\newcommand{\R}{\mathcal{R}}
\definecolor{red}{rgb}{0.8,0,0}
\definecolor{violet}{rgb}{0.4,0,0.4}
\definecolor{green}{rgb}{0,0.5,0.0}
\definecolor{navy}{rgb}{0.0,0.0,0.6}
\definecolor{orange}{rgb}{0.8,0.2,0.0}

\begin{document}
\title{\bf Superfluidity and Superconductivity in Neutron Stars}

\author{
      Brynmor~Haskell   \\
              {\small\it Nicolaus Copernicus Astronomical Center, Polish Academy of Sciences,}\\
{\small\it Bartycka 18, 00-716, Warszawa, Poland\vspace{0.6cm}}\\
            \and
       Armen~Sedrakian \\
   {\small\it     Frankfurt Institute for Advanced Studies,
    Ruth-Moufang-str. 1, D-60438 Frankfurt-Main, Germany}
}
\date{\small\today}

\maketitle

\begin{abstract}
  This review focuses on applications of the ideas of superfluidity
  and superconductivity in neutron stars in a broader context, ranging
  from the microphysics of pairing in nucleonic superfluids to macroscopic
  manifestations of superfluidity in pulsars. The exposition of the basics of
  pairing, vorticity and mutual friction can serve as an introduction
  to the subject. We also review some topics of recent interest,
  including the various types of pinning of vortices, glitches, and
  oscillations in neutron stars containing superfluid phases of
  baryonic matter.
\end{abstract}

\section{Introduction}
\label{sec:intro}

Neutron stars are one of the most extreme astrophysical laboratories
in the universe. They allow us to probe physics in strong gravitational
fields in the regime where general-relativistic corrections can be as 
large as  $ 20\%$, the magnetic fields deduced at their surfaces
$B\le 10^{15}$ G are the largest measured in Nature, and their
interiors are expected to contain the densest forms of matter. For
typical neutron star masses $\simeq 1.4-2.0\ M_{\odot}$ ($M_{\odot}$
being the solar mass) and radii $R\simeq 10-14$ km the central
densities of neutron stars can easily exceed the nuclear saturation
density $n_s=0.16$ fm$^{-3}$ by factors of a few up to ten. 

At the same time neutron stars are extreme low-temperature
laboratories: the high densities of their interiors imply large Fermi
energies of fermions $\ep_F\simeq 10-100$ MeV, which turn out to be
much higher than the characteristic interior temperatures of mature
neutron stars $T\sim 10^8$~K $\simeq 0.01$ MeV.  Because of the
attractive long-range component of the nuclear force and the high
degeneracy $\ep_F \gg T$ the neutrons and protons (and presumably some
hyperons) become superfluid and superconducting at critical
temperatures of the order of $T_c\simeq 10^9$~K.

Neutron superfluidity in a neutron star crust and its core, as well
as proton superconductivity in the core, profoundly alter its
dynamics, just as the emergence of these phenomena does in terrestrial
experiments. For example, superfluid neutrons can now flow
relative to the `normal' component of the star with little or no
viscosity, as standard reactions and scattering processes giving rise
to bulk and shear viscosity are strongly suppressed. An important
factor in neutron star dynamics is the appearance of an array of
vortices in the neutron condensate. In analogy to a laboratory superfluid in
a rotating container, the neutron superfluid mimics large scale rotation
by creating an array of quantised vortices each carrying a quantum of
circulation. Interactions between vortices and the normal component
open a new dissipative channel, known as mutual friction. These
interactions may be strong enough to `pin' the vortices and freeze the
rotation rate of the superfluid neutrons. The angular momentum thus
stored is then released catastrophically during discrete events, which
are thought to be the cause of the observed `glitches', i.e., sudden
spin-up episodes observed in pulsars. These phenomena reflect 
the interior dynamics of neutron stars and thus can potentially
provide an insight into the physics of superfluids in their
interiors.

This chapter provides an educational introduction and an overview of
the field of superfluidity and superconductivity in neutron stars.
The first part of the chapter reviews the microphysics of nuclear
pairing in neutron stars by providing an elementary introduction to
the microscopic theory of nuclear pairing and a review of current
issues such as medium polarization corrections to the pairing,
pairing in higher partial waves and in strong magnetic fields
(Sec.~\ref{sec:micro}). The interaction of vortices with the ambient
fluid at the microphysical level, which leads to the phenomenon of 
mutual friction between the superfluid and the normal fluid, is
reviewed in Sec.~\ref{sec:mutial_friction}.  This is followed by a
discussion of hydrodynamics of superfluids in neutron stars
(Sec.~\ref{sec:superfluid_hydro}). Section \ref{sec:pinning} is devoted to
the interactions of vortices with flux-tubes and nuclear clusters,
i.e., their pinning to various structures. This is followed by a
discussion of the macrophysics of rotational anomalies in neutron
stars in Sec.~\ref{sec:macro_super}. We provide our concluding remarks
in Sec.~\ref{sec:conclusions}.
 
\section{Microscopic pairing patterns in neutron stars} 
\label{sec:micro}

\subsection{General ideas}

The microscopic understanding of the pairing mechanism in nucleonic
matter in neutron stars is based on the theory advanced by Bardeen,
Cooper, and Schrieffer (BCS) in 1957 to explain the superconductivity
of some metals at low temperatures~\cite{1957PhRv..108.1175B}. The key
ingredient of this theory is the notion of an attractive interaction
between two electrons which is mediated by lattice phonon
exchange. According to the Cooper theorem \cite{1956PhRv..104.1189C}
low-temperature fermions which fill a Fermi sphere can bind to form
Cooper pairs if there is an attractive interaction between them. The
bound states of electrons (typically with total spin 0) form a
coherent many-body state which carries an electric current without any
resistivity below a certain critical temperature $T_c$. The
overwhelming success of the BCS theory in explaining the wealth of
experimental data encouraged applications of the key ideas of this
theory in other fields of physics, including nuclear physics. In
contrast to electronic materials, where the direct interaction between the
electrons is repulsive due to the Coulomb force between same-charge
particles, in nuclear systems the dominant long-range piece of the
interaction between the nucleons (neutrons and protons) is
attractive. It is not surprising then that superconductivity and
superfluidity in nuclear systems - finite nuclei and neutron stars -
were conjectured shortly after the advent of the BCS theory by A. Bohr
et al.~\cite{1958PhRv..110..936B}, Migdal~\cite{1959NucPh..13..655M} and others. 

Fully microscopic calculations of the pairing properties of neutron
and proton matter in neutron stars were carried out following the
discovery of pulsars in 1967 and their identification with neutron
stars. Although at the time nuclear interactions were not known as
precisely as nowadays, the first computations of the pairing gaps of
about 1 MeV in neutron and proton matter are consistent with present
day calculations~(see the
reviews~\cite{2001LNP...578...30L,2003RvMP...75..607D,2006pfsb.book..135S,2013arXiv1302.6626P,2014arXiv1406.6109G}
and references therein).

A useful reference for the understanding of the patterns of pairing in
neutron stars is the partial wave analysis of the nuclear
interaction. In fact, the experimental measurements of the nuclear
scattering are given per partial wave. At low energies the
nucleon-nucleon ($nn$) scattering is dominated by two $S$-wave
interactions, specifically the $^3S_1$--$^3D_1$ coupled partial wave
and the $^1S_0$ partial wave; here we use the standard spectroscopic
notations to specify the scattering channels, i.e., $^{2S+1}L_J$, with
$L=0,1,2$ mapped to $S$, $P$, $D$, where $L$ is the orbital angular
momentum, $S$ is the total spin and $J$ is the total angular momentum,
which is the sum of the former two vectors. Thus, at low energies the
$L=0$ states, which have symmetrical wave-function in the coordinate
space, dominate.  The total wave function contains however spin and
isospin components which must be selected in a manner to satisfy the
Pauli principle, which dictates that the total wave function must be
anti-symmetrical. As a consequence, the neutron-neutron and
proton-proton scattering which always has a total isospin $T=1$
symmetrical component, cannot occur in the spin symmetrical $S=1$ and
spatially symmetrical $L=0$ state. Therefore, the strongly
attractive interaction in the $^3S_1$--$^3D_1$ channel which binds the
deuteron (binding energy $E_{d}=-2.2$ MeV) does not lead to pairing in
neutron dominated matter, where neutrons and protons form vastly
separate Fermi surfaces. Then, for large differences in the numbers
of protons and neutrons (isospin asymmetry) only same isospin Cooper
pairs can arise in the remaining $^1S_0$ partial wave channel.

\begin{figure}[t] 
\begin{center}
\includegraphics[width=10.0cm,keepaspectratio]{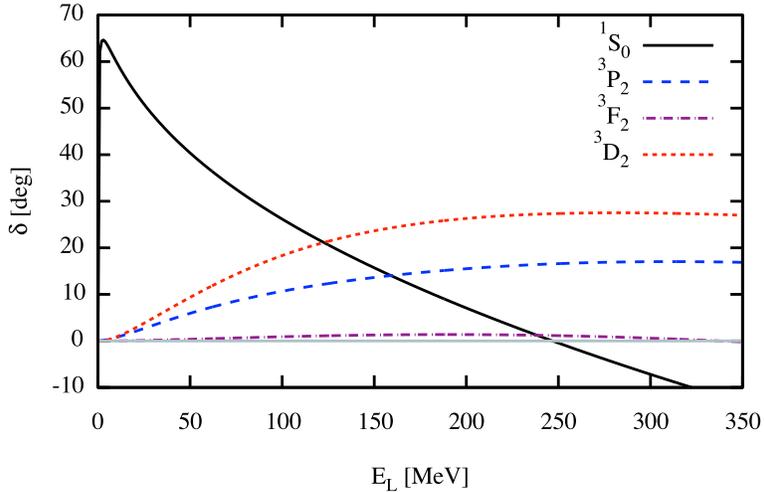}
\caption{The nucleon-nucleon scattering phase shifts as a function of
  laboratory energy for the channels where pairing in neutron stars
  matter appears. The $S$ and $P$ wave scattering is responsible for
  neutron-neutron and proton-proton pairing, whereas the $^3D_2$ wave
  scattering can occur only between neutrons and protons. The $P$ and
  $F$ waves are coupled by the non-central tensor component of the nuclear
  interaction. }
\label{Pairing_fig:1} 
\end{center}
\end{figure}

At laboratory energies of $nn$ scattering larger that $E_L=250$ MeV
the measured scattering phase-shift in the $^1S_0$-wave interaction
channel becomes negative, i.e., the interaction becomes repulsive, see
Fig.~\ref{Pairing_fig:1}. However, already at $E_L\simeq 160$ MeV the
$^3P_2-^3F_2$ tensor interaction becomes the most attractive channel
for $T=1$ (neutron-neutron and proton-proton) pairs.  The
corresponding density in neutron star matter is obtained by noting
that the center of mass energy of two scattering nucleons is $E_L/2$,
which should be of the order of the Fermi energy of neutrons or
protons. (Here we specialize the discussion to the high-density and
low-temperature regime of interest to superfluidity in neutron stars).
Neutron Fermi energies become of the order of $\ep_{Fn}\simeq 60$~MeV
at the nuclear saturation density $n_s=0.16$ fm$^{-3}$. Thus, we
anticipate that neutron pairing in the $^1S_0$-wave vanishes at
densities slightly above the saturation density and that the core of the
star contains  superfluid featuring neutron pairs in the
$^3P_2$--$^3F_2$ partial wave. The spatial component of the
wave-function of these Cooper pairs is anti-symmetrical whereas the
spin ($S=1$) and isospin ($T=1$) components are symmetrical. Clearly,
the pairing in this so-called triplet spin-1 channel is consistent
with the Pauli principle for two neutrons. Because the proton fraction
in a neutron star core is small, about $5$-$10\%$ of the net number
density, their Fermi energies, and consequently the center of mass
scattering energies, remain low. Therefore, proton pairs arise in the
$^1S_0$-wave up to quite high densities. It is conceivable that at
densities higher than a few times the nuclear saturation density
higher partial waves can contribute to the pairing in neutron star
matter. For example, if the partial densities of neutrons and protons
are forced to be close to each other by some mechanism, i.e., matter
is isospin symmetrical, then neutron-proton pairs can be formed in the
most attractive $^3D_2$ partial wave with a wave function which is
symmetrical in space, antisymmetrical in isospace ($T=0$) and
symmetrical in spin ($S=1$).  A mechanism that can enforce equal
numbers of neutrons and protons is meson
condensation~\cite{weber_book}.

The BCS theory was originally formulated in terms of a variational
wave function of a coherent state which minimized the energy of an
ensamble of electrons interacting via contact (attractive)
interaction~\cite{1957PhRv..108.1175B}. Here we will outline an
alternative formulation based on the method of {\it canonical
transformations} due to Bogolyubov~\cite{Bogoljubov1958}.

Consider a macroscopic number of $N$ fermions which are
described by the {\it pairing Hamiltonian} 
\bea\label{eq:BCS_Hamiltonian}
H-\mu N = \sum_{\vecp,\sigma}\ep_{\vecp} 
a^{\dagger}_{\vecp,\sigma}a_{\vecp,\sigma}
- \sum_{\vecp_1+\vecp_2=\vecp_3+\vecp_4}
\hspace{-0.7cm}V_{\rm eff}(\vecp_1,\vecp_2;\vecp_3,\vecp_4)
a^{\dagger}_{\vecp_3}a^{\dagger}_{\vecp_4} a_{\vecp_1} a_{\vecp_2}.
\eea 
The first term is the kinetic energy, the second term is the
attractive interaction energy, where
$V_{\rm eff}(\vecp_1,\vecp_2;\vecp_3,\vecp_4)$ is an effective pairing
interaction. Here $a^{\dagger}_{\vecp,\sigma}$ and $a_{\vecp,\sigma}$
are the particle creation and annihilation operators for particles
with spin $\sigma = \uparrow\downarrow$ and momentum $\vecp$. Note
that we work in the grand-canonical ensemble, so that instead of
fixing the number of particles we assume that our system is connected
to a reservoir of particles; $\mu$ is the chemical potential - the
energy needed to add or remove a particle to the system. The
application of the method of canonical transformations to the
Hamiltonian \eqref{eq:BCS_Hamiltonian} requires new creation and
annihilation operators defined as
\bea && a_{\vecp,\uparrow}
= u_p\alpha_{\vecp\uparrow}+v_p\alpha^{\dagger}_{-\vecp\downarrow},\\
&& a_{\vecp,\downarrow} =
u_p\alpha_{\vecp\downarrow}-v_p\alpha^{\dagger}_{-\vecp\uparrow}.
\eea 
The requirement that the anti-commutation relations obeyed by the
new operators are the same as those obeyed by the original fermionic 
ones leads us to 
\be
\{\alpha_{\vecp,\sigma},\alpha^{\dagger}_{\vecp',\sigma'}\} =
\alpha^{\dagger}_{\vecp,\sigma}\alpha_{\vecp',\sigma'}
+\alpha_{\vecp',\sigma'}\alpha^{\dagger}_{\vecp,\sigma} =
\delta_{\vecp\vecp'}\delta_{\sigma,\sigma'}.  
\ee 
It follows then that the functions $u_p$ and $v_p$ are not
independent, but $ u_p^2+v_p^2 = 1$, i.e., there is a single
{\it independent} function, say $v_p$. 
This parameter is found from the  minimization of 
the statistical average of the Hamiltonian \eqref{eq:BCS_Hamiltonian}
\be 
E-\mu N = \langle H-\mu N \rangle  ,
\ee 
where 
$\langle\dots\rangle $
stands for mean value with the occupation numbers defined as
$\langle
\alpha^{\dagger}_{\vecp,\downarrow}\alpha_{\vecp,\downarrow}\rangle =
n_{\vecp,\downarrow}$
and
$ \langle
\alpha^{\dagger}_{\vecp,\uparrow}\alpha_{\vecp,\uparrow}\rangle =
n_{\vecp,\uparrow}.  $
The energy is then given by 
\bea \label{eq:BCS_energy} E-\mu N &=&
\sum_{\vecp} \ep(\vecp) \left[
  u_p^2(n_{\vecp,\uparrow}+n_{\vecp,\downarrow})+
  v_p^2(2-n_{\vecp,\uparrow}-n_{\vecp,\downarrow})
\right] \nonumber\\
&-& \sum_{\vecp\vecp'}V_{\rm eff}(\vecp,\vecp') u_pv_p u_{p'}v_{p'} Q(\vecp)
Q(\vecp') , 
\eea 
where
$Q(\vecp) \equiv (1-n_{\vecp,\uparrow}-n_{\vecp,\downarrow})$.
Eliminating $u_p$ from \eqref{eq:BCS_energy} via $u_p^2=1-v_p^2$ and
performing variations with respect to $v_p$ we find then
\be \label{eq:BCS_minimum} 2\ep_p =\frac{\Delta(\vecp)
  (1-2v_p^2)}{u_pv_p}, 
\ee
where 
\be\label{eq:Gap_equation} \Delta (\vecp) =
\sum_{\vecp'}V_{\rm eff}(\vecp,\vecp') u_pv_p (1-n_{\vecp,\downarrow} -
n_{\vecp,\uparrow}), 
\ee 
is the so-called {\it gap equation}. From
Eq.~\eqref{eq:BCS_minimum} we find for the {\it Bogolyubov
  amplitudes} 
\be u_p^2 = \frac{1}{2}
\left(1+\frac{\ep_p}{E_p}\right),\quad\quad v_p^2 = \frac{1}{2}
\left(1-\frac{\ep_p}{E_p}\right), 
\ee 
where the quasiparticle spectrum
in the superconductor is defined as 
\be \label{eq:BCS_spectrum} E_p = \sqrt{\ep_p^2+\Delta_p^2}.  
\ee 
On substituting $u_pv_p = \Delta(\vecp)/2E_p$ in
Eq.~\eqref{eq:Gap_equation}, we find a non-linear integral equation
for the gap function $\Delta(\vecp)$ which can be solved for any given
effective interaction $V_{\rm eff}(\vecp,\vecp')$. It is
easy to verify that $E_p$ is indeed the quasiparticle energy, by taking
the variation of the total energy with respect to the occupation
numbers, i.e., by computing the variation\footnote{When
  computing the variation with respect to the occupation numbers one
  needs to assume that the Bogolyubov coefficients are constant,
  because they are determined from the condition
  $ \delta (E-\mu N)/\delta v_p = 0.  $ }
\be \label{eq:BCS_spectrum2} \frac{\delta (E-\mu N)}{\delta
  n_{\vecp,\uparrow}} = \ep_p(u_p^2-v_p^2) + 2 u_pv_p\Delta(\vecp) =
E_p.  \ee 
Eq.~\eqref{eq:BCS_spectrum} demonstrates the fundamental property of
the superconductors: {\it the spectrum of a superconductor contains an
  energy gap $\Delta$.}  As a consequence the excitations can be
created in the system if a Cooper pair breaks, which means that energy
of the order of $2\Delta$ must be supplied to the superconductor.  The
main property of superconductors - the absence of dissipation of
current - follows from the existence of the gap in their spectrum. In
the case of uncharged fermionic superfluids (e.g. neutron matter) the
same property is referred to as superfluidity (fluid motions without
dissipation).  In equilibrium, the occupation numbers of fermions are
given by the Fermi function $f(p) = (e^{E_p/T}+1)^{-1}$, where $T$ is
the temperature. At low temperatures the fermionic momenta are
restricted to the vicinity of the Fermi surface; then, assuming an
isotropic ($S$-wave) interaction, we can simplify the gap equation by
changing the integration measure
$\sum_{\vecp} = m^*p_F\int d\ep_p\int d\Omega $ to find
\be 
1 = G\nu \int_0^{\Lambda} \frac{d\ep_p}{2\sqrt{\ep_p^2+\Delta^2}}
\tanh \left(\frac{\sqrt{\ep_p^2+\Delta^2}}{2T}\right), 
\quad
\ee
where $\nu = p_Fm^*/\pi^2$ is the {\it density of states}, $m^*$ is
the effective mass, $p_F$ is the Fermi momentum, and for the sake of
illustrations below we assume a momentum-independent contact interaction
$V_{\rm eff}(\vecp,\vecp') =  G$, which in turn requires  a cut-off
$\Lambda$ to regularize the integral in the ultraviolet. The latter
cut-off is physically well-motivated, as the effective pairing
interactions are typically localized close to the Fermi
surface.\footnote{The full nuclear interaction can be renormalized via
  resummations of infinite series such as to contain only components
  close to the Fermi surface. }

Consider now  analytical solutions of the gap equation in the limiting cases 
$T\to T_c$ and $T\to 0$, where $T_c$ is the critical temperature of
phase transition.  For $T =  0$, the $\tanh$ function is unity and 
a straightforward integration gives 
\be 
1 = \frac{G\nu}{2}{\rm arcsinh}\left(\frac{\Lambda}{\Delta(0)}\right) \simeq 
\ln \left(\frac{2\Lambda}{\Delta(0)}\right),
\ee
where in the last step we assumed {\it weak coupling }, i.e., 
 $\Delta\ll \Lambda$ to expand $\displaystyle{\lim_{x \to \infty}} {\rm arcsinh} ~x =
 \ln(2x) + O(x^2)$;  this can be written in a more familiar form
\be\label{eq:BCS_contact}
\Delta(0) = 2\Lambda \exp \left(-\frac{2}{G\nu}\right), 
\ee
which is the famous {\it gap equation} of the BCS theory. It
demonstrates the exponential sensitivity of the pairing gap to the
effective  attractive interaction $G$. In the limit $T\to T_c$, 
we can set in the integrand of the gap equation
$\Delta = 0$. An elementary integration then gives 
$ T_c =
(2\Lambda\gamma/\pi)\exp\left(-2/G\nu\right),
$
where $\gamma \equiv e^C$ and $C = 0.57$ is the Euler constant.
Combining the results for $T_c$ and $\Delta(0)$ we obtain a relation
between them: $\Delta(0) = \pi T_c/\gamma= 1.76 ~T_c.$ 

The limiting expressions for the gap function can be easily extended to
include the next-to-leading order terms in the two limiting cases discussed above:
\be \Delta(T)
= \Delta(0) - \sqrt{2\pi\Delta(0) T}
\exp\left(-\frac{\Delta(0)}{T}\right) 
\ee 
for $T\to 0$ and 
\be \Delta(T) =
\pi\sqrt{\frac{8}{7\zeta(3)}}\left[T_c(T_c-T)\right]^{1/2} =
3.06\left[T_c(T_c-T)\right]^{1/2}
\ee 
for $T\to T_c$, where $\zeta(x)$ is the Riemann's $\zeta$-function
with $\zeta(3)= 1.20205$. The temperature dependence of the gap
function in the whole temperature regime can be obtained numerically
and analytical fits, useful for practical applications, are given in
Ref.~\cite{1959ZPhy..155..313M}. Having the temperature dependence of
the gap one still needs fit formulae to the gap function at zero
temperature $\Delta(0)$ (or equivalently the critical temperature
$T_c$) as a function of density or Fermi momentum.  We point out
that accurate fits can be obtained with the functional form
\bea
\Delta (k_F) = a \exp(-k_F^2) + \sum_{n=0}^4 c_n k_F^n,
\eea
where the fit coefficients to the $S$- and $P$-wave neutron and
$S$-wave proton gaps can be found in Ref.~\cite{2015PhRvC..91c5805S}.

\subsection{Effective interactions}

An important issue in computations of gaps in neutron star matter
is the proper determination of the effective pairing interaction
$V_{\rm eff}(\vecp,\vecp')$. As a first approximation one may use the
bare nuclear interaction in the gap equation, which provides us with a
useful reference result. The most important correction to this
interaction arises from {\it polarization effects} or {\it screening},
which in diagrammatic language can be understood as a summation of
infinite series of particle-hole diagrams. We will review these
effects on the basis of the {\it Landau Fermi liquid theory} and will
compare to the alternatives thereafter. In our discussion we will
follow the original Landau approach; for computations which are based
on this type approach, but include more advanced diagrammatic
treatments of the problem see
Refs.~\cite{1993NuPhA.555..128W,1996PhLB..375....1S,2003NuPhA.713..191S,2004NuPhA.731..392L,2005PhRvC..71e4301S}.

We now suppose that the bare interaction depends only on the momentum
transfer in the collision $\vecq = \vecp_1-\vecp_3$ and write it in
terms of all possible combinations of the spin and isospin components:
\be \label{PH2}
V_{\rm bare}(\vecq) = \frac{1}{\nu}\left\{
F_{\vecq} + G_{\vecq} (\vecsigma \cdot \vecsigma')
          + \left[F'_{\vecq} + G'_{\vecq} (\vecsigma \cdot \vecsigma')
            \right](\vectau\cdot \vectau')\right\},
\ee
where $\vecsigma$ and $\vectau$ are the vectors formed from the Pauli
matrices in the spin and isospin spaces and
$F_\vecq,F'_\vecq, G_\vecq$ and $G'_\vecq$ are the so-called {\it
  Landau parameters}.  For simplicity the tensor and spin-orbit
interactions are neglected as they are small in neutron matter.  The
summation of geometrical series of particle-hole diagrams then gives
for the effective interaction
\bea \label{eq:Landau}
 V_{\rm eff} =\frac{1}{\nu}\Big\{ \tilde F_{\vecq}
                   + \tilde G_{\vecq} (\vecsigma \cdot \vecsigma')
                   + \left[\tilde F'_{\vecq}
                   + \tilde G'_{\vecq}
                     (\vecsigma \cdot \vecsigma')
\right](\vectau\cdot \vectau')\Big\},\nonumber\\
\eea
where $\tilde A_{\vecq} = A_{\vecq}[1+\Lambda(q)A_{\vecq}]^{-1} $, $A$
stands for any of the Landau parameters. It is seen that the screening
leads to a renormalization of the Landau parameters $A\to \tilde A$,
where the function $\Lambda(q)$, which represents the single
particle-hole loop, is the {\it Lindhard function.} In the
low-temperature regime of interest the momenta of fermions are
restricted to lie on their Fermi surface, therefore the Landau
parameters will depend only on the relative orientation of the momenta
of particles. Then, it is useful to expand these into spherical
harmonics
\be \label{eq:Landau_spherical} A(q) =\sum_l A_l P_l(\cos \theta)\,,
\ee 
where $P_l$ are the Legendre polynomials, $A$ stands for any of the
Landau parameters above, $\theta$ is the scattering angle which is
related to the magnitude of the momentum transfer via
$q = 2p_F \sin\theta/2$, where $p_F$ is the Fermi momentum. 

In pure neutron matter ($\vectau \cdot \vectau' = 1$) keeping the
lowest-order harmonics in the expansion (\ref{eq:Landau_spherical})
and for scattering with total $S = 0$ (i.e.
$\vecsigma\cdot\vecsigma' = -3$) the effective pairing interaction is
given by
\be \label{eq:effective_int_landau} \nu(p_F) V_{\rm eff}(q) =
F_0^{n}\left[1- {\cal L}(q) F_0^{n}\right]
-3G_0^{n}\left[1- {\cal M}(q)    {G}^{n}\right].  
\ee 
where $F^{n} = F + F'$ and $G^{n} = G + G'$ describe the effective
interaction in the density and spin channels respectively,
${\cal L}(q) = \Lambda(q)[1+\Lambda(q)F_0^{n}]^{-1}$ and
${\cal M}(q) = \Lambda(q)[1+\Lambda(q)G_0^{n}]^{-1}$ are screening
corrections to the direct part of the effective interaction
$\propto 1$. In general the Lindhard function is complex; in the
low-temperature regime of interest its imaginary part (which is
related to the damping of particle-hole excitations) can be
neglected. It assumes a simple form for zero energy transfer at fixed
momentum:
$\Lambda (x)= -1+ (2x)^{-1} (1-x^2) \ln \vert (1-x)/(1+x)\vert$, with
$x = q/2p_F$~\cite{Fetter}.

The Landau parameters and the effective interaction in
Eq.~\eqref{eq:effective_int_landau} have been computed extensively 
over the past several decades within various approximations. There is
a general agreement that the density fluctuations
$\propto F^n_0$ enhance, whereas the spin-fluctuations $\propto G^n_0$
reduce the attraction in the pairing interaction. The overall effect
of the spin fluctuations at subnuclear densities is numerically larger
and, consequently, fluctuations reduce the pairing gap in neutron
matter. 
\begin{figure}[t] 
\begin{center}
\includegraphics[width=10.0cm,keepaspectratio]{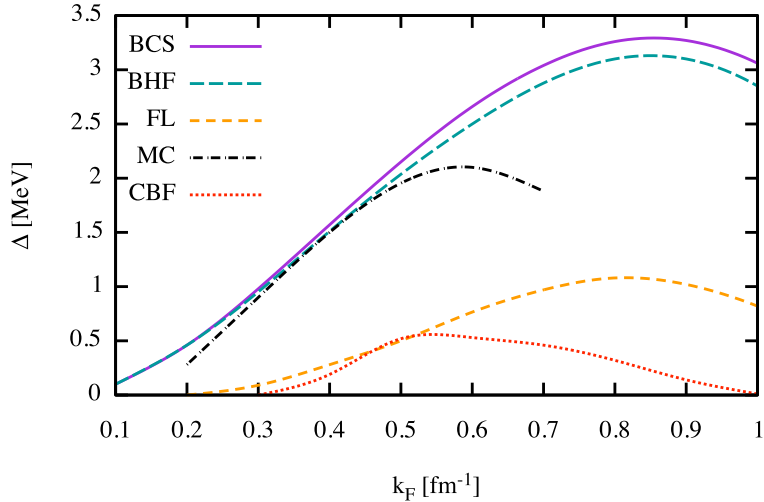}
\caption{The pairing gap in the low-density neutron matter relevant
  for neutron star crusts as a function of Fermi momentum. The curves
  labeled ``BCS'' and ``BHF'' show the result for low-momentum
  interaction using bare and medium modified single particle spectra,
  whereby the renormalization of the particle spectrum is taken into
  account in the Bruckner-Hartree-Fock
  theory~\cite{2003PhLB..576...68S}. The screening effects, which
  strongly suppress the gap, are shown on the examples based on
  Fermi-liquid ``FL''~\cite{1993NuPhA.555..128W} and correlated basis
  functions ``CBF'' theories~\cite{1993NuPhA.555...59C}. We also show
  the results of auxiliary field Monte Carlo ``MC''
  simulations~\cite{2008PhRvL.101m2501G}, which are closer to the
  results without screening corrections.  }
\label{Pairing_fig:2} 
\end{center}
\end{figure}
As can be seen from Eq.~\eqref{eq:BCS_contact} the dimensionless
quantity (coupling) determining the magnitude of the gap involves,
apart from the effective interaction, also the density of states. In
general the quasiparticle spectrum can be written as
\bea\label{eq:varep}
\ep(p) = \frac{p^2}{2m}+ U(p),
\eea 
where $U(p)$ is the single particle potential felt by a particle in
the nuclear environment. The full momentum dependence in Eq.~\eqref{eq:varep} 
 can be approximated by introducing
an effective quasiparticle mass. An
expansion of the potential $U(p)$ around the Fermi momentum leads to
\be
\ep(p) = \frac{p_F}{m^*}(p-p_F) - \mu^*, \quad
\frac{m^*}{m}  = \left(1+\frac{m}{p_F}
\frac{\partial U(p)}{\partial p}\Big\vert_{p=p_F}\right)^{-1}, 
\ee
where $\mu^*\equiv -\epsilon(p_F)+\mu - U(p_F)$. Because the effective
mass $m^*/m< 1$ in neutron matter, the modifications of the
single-particle energies lead to a reduction of the gap and will also
have a significant effect on the multifluid hydrodynamics of the
system, as discussed in Sec.~\ref{sec:superfluid_hydro}.

Correlations in neutron matter can be studied in a number of
alternative theories, for example, Monte Carlo sampling of systems
with odd and even numbers of neutrons, whereby the gap is
defined as the energy difference between odd and even
states~\cite{2008PhRvL.101m2501G,2010PhRvC..81b5803G}. Wave-function
based approches minimize the energy of the BCS state using correlated
basis functions (CBF), which are built to account for the operator
structure of the nuclear interaction~\cite{1993NuPhA.555...59C}.

In Fig~\ref{Pairing_fig:2} we show selected benchmark results for
pairing gaps in different theories of pairing. The BCS result with
bare single particle energies was obtained using a low-momentum
interaction, which leads to slightly larger gaps compared to full
(realistic) interaction which contains a hard core. The modifications of
the single particle spectrum, computed within the
Bruckner-Hartree-Fock theory (without invoking effective mass
approximation) leads to a moderate reduction of the
gap~\cite{2003PhLB..576...68S}. If one includes the screening
corrections in the effective interaction, then the gap is reduced by a
factor of three. There is consensus among the
Fermi-liquid~\cite{1993NuPhA.555..128W,1996PhLB..375....1S,2003NuPhA.713..191S,2004NuPhA.731..392L,2005PhRvC..71e4301S}
and CBF approaches~\cite{1993NuPhA.555...59C,2016arXiv161202188P} on
the magnitude of the reduction.  Auxiliary field Monte Carlo
calculations on the other hand predict pairing gaps at low densities
which are much closer to the BCS result~\cite{2008PhRvL.101m2501G}.

\subsection{Higher partial wave pairing}

As seen from Fig.~\ref{Pairing_fig:1} at high densities (energies) the dominant
pairing interaction between neutrons is in the $^3P_2-^3F_2$ partial
wave. The pairing pattern for spin-1 condensates is more complex than
for spin-0 $S$-wave condensates because of competition between states
with various projections of the orbital angular momentum and
complications due to the tensor coupling of the $^3P_2$ partial wave
to the $^3F_2$ one, see
Refs.~\cite{1998PhRvC..58.1921B,2003NuPhA.720...20Z,2004PhRvL..92h2501S,2014PhRvC..90d4003M}.
To solve the $P$-wave gap equation (\ref{eq:Gap_equation}) one starts
with an expansion of the pairing interaction in partial waves
\be
 V_{\rm eff.} (\vecp,\vecp') =
  4\pi \sum_L (2L+1) P_L({\hat{\vecp}\cdot\hat{\vecp}'}) V_L(p,p'),
\ee
where $P_L$ are the Legendre polynomial, and an associated 
expansion of the gap function in spherical harmonics $Y_{LM}$
\begin{equation}
  \Delta(\vecp) =
  \sum_{L,M} \sqrt{4\pi\over 2l+1} Y_{LM}(\hat \vecp) \Delta_{LM}(p),
\end{equation}
where $L$ and $M$ are the total orbital momentum and its
$z$-component.  The non-linearity of the gap equation prevents a
straightforward solution for its components $\Delta_{LM}(p)$; a common
approximation is to perform an angle average in the denominator of the
kernel of the gap equation by replacing the factor
$\sqrt{\ep(p)^2+|\Delta(\vecp)|^2}\to
\sqrt{\ep(p)^2+D(p)^2}$
where the {\it the angle averaged gap} is given by 
\begin{equation}
  D(p)^2 \equiv\int \frac{d\Omega }{4\pi} \, \vert \Delta(\vecp)\vert ^2 =
  \sum_{L,M} {1\over 2L+1} |\Delta_{LM}(k)|^2 .
\end{equation}
With this approximation the angular integrals are trivial and we
obtain a one-dimensional gap equation for the $L$-th component of the gap 
\begin{equation}
  \Delta_L(p) = - \int_0^\infty
\frac{dp' p'}{\pi} \frac{V_L(p,p')}{\sqrt{ \ep(p')^2 +
  \left[\sum_{L'} \Delta_{L'}(p')^2 \right] }}
  \Delta_L(p').
\end{equation}
Although the denominator contains a sum over gap components with different
values of $L$, the contributions from channels other than the
$^3P_2$ can be neglected as they are numerically insignificant in the
range of densities (energies) where $P$-wave paring is dominant.  

The non-central tensor part of the nuclear force couples the $^3P_2$
wave to the $^3F_2 $ wave; this coupling affects the value of the
gap. The modification of the gap equation which takes into account
this tensor coupling requires a simple doubling of the components of
the gap equation.  The coupled channel gap equation reads
\bea
 \left( \begin{array}{c} \Delta_L \\ \Delta_{L'} \end{array} \right)
=  \int_0^\infty\!\!\! \frac{dp' p'^2}{\pi E(p')} 
 \left( \begin{array}{rr}
  -V_{LL} & V_{LL'} \\ V_{L'L} & -V_{L'L'} 
 \end{array} \right)
 \left(\begin{array}{c} \Delta_L \\ \Delta_{L'} \end{array}\right),
\eea
where now
$ D(k)^2 = \Delta_{L}(k)^2 + \Delta_{L'}(k)^2$.~\footnote{Note that
  the tensor coupling arises also in the case of pairing in the
  $^3S_1$-$^3D_1$ channel which acts only between neutrons and protons
  and is relevant when the isospin asymmetry between neutrons and
  protons in not too
  large~\cite{2001PhRvC..64f4314L}.} 

Note that the angle averaged approximation provides a numerically
accurate value of the angle averaged gap on the Fermi
surface. However, in a number of problems, such as neutrino and axion
emission from $P$-wave superfluids the angle dependence of the gap
equation is an important factor and should be taken into account by
solving the gap equation without the angle averaged approximation. Such
solutions show that the angle dependence of the gap function can leads
to nodes on the Fermi surface, as for example in the case of solutions
of the form $\Delta(\theta) = \Delta_0 \sin \theta$. However,
``stretched'' solutions with fully gaped Fermi surface
$\Delta (\theta) = \Delta_0 (1+ \cos^2 \theta)$ are viable candidates
for angle dependence of the pairing gap and it is difficult to
distinguish between these options from energy minimization arguments
alone.

Figure \ref{Pairing_fig:1} shows that the interaction is attractive
among neutrons and protons in the $^3D_2$ channel and it is stronger
than the attraction in the $^3P_2$ channel. Thus, in a hypothetical high-density isospin symmetrical form of nuclear matter one would expect
$D$-wave pairing instead of the $P$-wave pairing, which is the
dominant channel in the high density neutron matter. Consequently,
there should be a transition from $D$- to $P$-wave pairing as the
imbalance between neutrons and protons (isospin asymmetry) changes
from zero to larger values. Computations show that already small
asymmetries destroy the $D$-wave pairing in nuclear
matter~\cite{1996NuPhA.604..491A}, therefore for its realization one
needs nearly symmetrical nuclear matter. Above nuclear saturation
density such situations can arise in some special cases as, for
example, when a mesonic condensate forms~\cite{weber_book}.

\subsection{Effects of strong magnetic fields on pairing}
\label{sec:Pairing_Strong_B}

Neutron stars are highly magnetised objects and the magnetic field in the stellar
interior is modified by the presence of superconductivity. The
topology and properties of the magnetic field depend strongly on the
type of superconductivity, which depends on the Ginzburg-Landau 
parameter $\kappa_{\rm GL} = \lambda/\xi_\p$, where
$\xi_\p$ is the proton coherence length, which roughly determines the
size of the Cooper pairs, and $\lambda$ is 
the penetration depth of the magnetic field in
superconducting matter. In most of the neutron star core one has
$\kappa_{\rm GL}>1/\sqrt{2}$ and type II superconductivity is
expected, in which the magnetic field penetrates the superconductor by
forming an array of quantized flux tubes.  In laboratory type II
superconductors the field can only penetrate the superconductor for
field strengths between a lower critical field $H_{c1}$ and an upper
critical field $H_{c2}$. In neutron stars the situation is somewhat
different, as one still has the upper critical field $H_{c2}$
(essentially the field at which flux tubes are so densly packed that
their cores touch), but magnetic fields can still penetrate the core
below the lower critical field $H_{c1}$. This is the case because the
magnetic flux cannot be expelled effectively from the superconducting
core due to its high electric conductivity; the time-scale for such
expulsion is of the order of the secular timescales
\cite{1969Natur.224..872B}.  In the deep core of the neutron star, on the
other hand, it is possible to have $\kappa_{\rm GL}<1/\sqrt{2}$, and
in this case we expect type I superconductivity, in which fluxtubes
are not energetically favourable and the field is arranged in domains
of unpaired proton matter of much larger spatial
 dimensions~\cite{1973Ap......9..134B,1997MNRAS.290..203S,2003PhRvL..91j1101L,2004PhRvC..69e5803B,2005PhRvD..71h3003S}.

A class of neutron stars known as magnetars have surface magnetic
fields of the order of $10^{15}$~G and it has been conjectured that
their interior fields could be by several orders of magnitude
larger~\cite{2015RPPh...78k6901T}.  While modifications of the
equation of state of matter require fields which are close to the
limiting fields $10^{18}$~G compatible with gravitational stability,
pairing phenomena which occur near the Fermi surface are affected by
much lower fields of the order $B\sim 10^{16}-10^{17}$~G.  The
interaction energy of the magnetic field with the nucleon spin is
$\mu_NB$, where $\mu_N=e\hbar/2m_p$ is the nuclear magneton.  A
numerical estimate gives
$\mu_NB\simeq \pi (B/10^{18}~ \textrm{Gauss})$~MeV. Therefore, fields
of the order of $10^{16}$~G would substantially affect the pairing
with gaps of the order of 1 MeV via the spin--$B$-field interaction.
The interaction of the magnetic field with the neutron or proton spin
induces an imbalance in the number of spin-up and spin-down particles,
which implies that the Cooper pairing in the $S$ wave will be
suppressed. Indeed in this case the number of Cooper pairs will be
limited by the number of spin-down particles, with the excess spin-up
particles remaining unpaired~\cite{2016PhRvC..93a5802S}.  This {\it
  Pauli paramagnetic suppression} acts for both proton and neutron
condensates and the associated critical field is within the range
${H}^{\rm Para}_{c}\sim 10^{16}$-$10^{17}$~G. Note that similar field
arises in the condensed matter theory and is known as the
Chandrasekhar-Clogston field.

In the case of the proton condensate the upper critical field $H_{c2}$
turns out to be smaller than the field associated with the Pauli
paramagnetic ordering~\cite{2015PhRvC..91c5805S}, therefore the proton
condensate is destroyed for even lower magnetic fields
$H_{c2}\simeq 10^{15}$~G.  We have seen that the $S$-wave gaps and in
particular the proton gap depends substantially on the density. If the
magnetic field strength can be assumed approximately constant in the
core of a magnetar the size of the superconducting region will depend on
the magnitude of the field $B$ via the condition $B\le H_{c2}$.

The role of the magnetic field in the $P$-wave pairing is not well
understood from microscopic point of view, but because the pairs in
this case form spin-1 objects, the spin-magnetic field interaction
will not be destructive. The consequences of the suppression of the
nucleonic pairing on the phenomenology of magnetars are discussed
elsewhere~\cite{2017arXiv170100895S}.

\section{Microphysics of mutual friction}
\label{sec:mutial_friction}

In this section we concentrate on the interaction of vortex lines in
neutron stars with ambient components. The discussion includes both
the neutron vortex lines which are formed in response to the rotation
of the star and the magnetic flux-tubes which, as discussed above,
form if the proton superconductor is type-II. This interaction is
known under the general name {\it mutual friction} and appears
naturally in the superfluid hydrodynamics including vortices, which we
will discuss in the following Sec.~\ref{sec:superfluid_hydro}.  The
mutual friction is an important ingredient of the description of
superfluid dynamics as it determines the dynamical time-scales of
coupling of superfluid to normal (non-superfluid) matter and
rotational anomalies in neutron stars, such as glitches, time-noise,
precession etc., which are in part discussed in the subsequent
Sec.~\ref{sec:macro_super}. The mutual friction is quantified in terms
of dimensionfull drag parameter $\eta$ defined via force exerted by
ambient fluids on the vortex
$
\vecf_d = \eta (\vecv_v-\vecv_e),
$
where $\vecv_v$ is the vortex velocity, $\vecv_e$ is the velocity of
the normal component. In superfluids it is balanced by the  Magnus
(lifting) force 
$
\vecf_M = \rho_n [(\vecv_s-\vecv_v)\times \vecnu],
$
acting on a vortex with circulation vector $\vecnu $ placed in a
superfluid flow with velocity $\vecv_s$.  It is, therefore, convenient
to use the dimensionless drag-to-lift ratio
${\mathcal R} = \eta/\rho_n\kappa$, where $\rho_n$ is the mass density
of the superfluid (neutron) component and $\kappa = \pi\hbar /m_n$ is
the quantum of circulation, $m_n$ being neutron mass. For massless
vortices $\vecf_M+\vecf_d = 0$, which is known as {\it the force
  balance equation}.  We shall discuss these quantities in more detail
in the context of superfluid hydrodynamics in
Sec.~\ref{sec:superfluid_hydro}.

The neutron vortices which carry the angular momentum of neutron star
interiors arise in the $S$ and $P$ wave superfluids where the Cooper
pairs have total spin-0 and spin-1, as discussed in
Sec.~\ref{sec:micro}. A vortex in a neutral fermionic superfluid has a
core of the order of the coherence length $\xi$, where quasiparticle
pairing is quenched and, therefore, scattering centers are available
for interactions.  Consider first a vortex in a neutron superfluid
with an $S$-wave symmetry of the condensate.  The core of the vortex
contains fermionic states which are given by
\cite{1965PKM.....3..380C}
\be\label{13}
\left(\begin{array}{c}
u_{q_{\pp},\mu}(\vecr_{\perp})\\ v_{p_{\pp}, \mu}(\vecr_{\perp})\end{array}\right) 
= e^{ip_{\pp}z} \left(e^{i\theta(\mu-\frac{1}{2})}
~e^{i\theta(\mu+\frac{1}{2})}\right) \left(\begin{array}{c}
u'_{\mu}(r)\\ v'_{\mu}(r)\end{array}\right), 
\ee
where $r,\theta, z$ are cylindrical coordinates with the axis of
symmetry along the vortex circulation (here $\pp$ and $\perp$ are the
parallel and perpendicular to vortex components), and $\mu$ is the
azimuthal quantum number, which assumes half-integer positive
values. It is seen that the states are plain waves along the vortex
circulation, which are quantized in the orthogonal direction. The
radial part of the wave-function is given by
\be\label{14}
\left(\begin{array}{c}
u'_{\mu}(r)\\ v'_{\mu}(r)\end{array}\right)=2\left(\frac{2}{\pi p_{\perp}r}\right)^{1/2}
e^{-K(r)} \left(\begin{array}{c}
{\rm cos}\left(p_{\perp}r - \frac{\pi\mu}{2} \right) \\ 
{\rm sin}\left( p_{\perp}r - \frac{\pi\mu}{2} \right)\end{array}\right),
\ee
where $p_{\perp} = \sqrt{p^2-p_F^2}$, $p_F$ being the neutron Fermi momentum, 
and the function in the exponent is given by
\be \label{15}
K(r) = \frac{p_F }{\pi p_{\perp} \Delta_{n}} \int_0^r\Delta(r') dr'\simeq
\frac{p_F r}{\pi p_{\perp}\xi}\, \left(1 + \frac{\xi e^{-r/\xi}}{r}\right).
\ee
The eigenstates of neutrons in the core of a vortex are given by 
\be \label{eq:vortex_eigenstate}
\ep_{\mu}(p) 
 \simeq \frac{\pi\mu\Delta^2_{n}}{2\ep_{Fn}}\left( 1+\frac{p^2}{2p_F^2}\right),
\ee
where $\ep_{Fn}$ is the Fermi energy of neutrons, $\Delta_{n}$ is the
asymptotic value of the gap far from the vortex core and the second
equality holds up to the next-to-leading order in small quantity
$p/p_F$.

Electrons will couple to the core quasiparticles of the neutron vortex
via the interaction of the electron charge $e$ with the magnetic
moment of neutrons $\mu_n = -1.913\mu_N$, where $\mu_N = e\hbar/2m_p$
is the nuclear magneton~\cite{1971PhRvD...4.1589F}. The momentum
relaxation time scale for electrons off neutron vortices is given
by~\cite{1989ApJ...342..951B} 
\bea \label{eq:S_wave_relax}
\tau_{eV}[^1S_0] = \frac{1.6\times
  10^3}{\Omega_s} \frac{\Delta_n}{T}
\left(\frac{\ep_{Fe}}{\ep_{Fn}}\right)^2
\left(\frac{\ep_{Fn}}{2m_n}\right)^{1/2}
\exp\left(\frac{\ep^0_{1/2}}{T}\right),  
\eea 
where $\ep_{Fe}$ is the Fermi energy of electrons, $\Delta_n$ is the
$S$-wave neutron pairing gap, $\ep^0_{1/2}$ is given by
Eq.~\eqref{eq:vortex_eigenstate} with $\mu=1/2$ and $\Omega_s$ is the
superfluid's angular velocity.  The relaxation time is strongly
temperature dependent because of the Boltzmann exponential factor
involving the eigenstates of the vortex core quasiparticles. 


The vortex structure of the $P$-wave superfluid was studied by Sauls
et al.\ \cite{1982PhRvD..25..967S} using a tensor order parameter
$A_{\mu\nu}(\vecr)$, $\mu, \nu = 1,2,3$ which is traceless and
symmetric. It can be decomposed in cylindrical coordinates ($r,\phi,z$)
as
\bea
A_{\mu\nu}  = \frac{\Delta}{\sqrt{2}} e^{i\phi} 
\Big\{
[f_1 \hat r_{\mu}\hat r_{\nu}
+f_2 \hat\phi_{\mu}\hat\phi_{\nu}
-(f_1+f_2) \hat z_{\mu}\hat z_{\nu}
+ig ( r_{\mu}\hat\phi_{\nu}+ r_{\nu}\hat\phi_{\mu}
)]\Big\},
\eea
where where $g(r)$ and $f_{1,2}(r)$ are the radial functions
describing the vortex profile and $\Delta$ is the average value of the
gap in the $^3P_2$ channel.  Vortices in a $P$-wave superfluid are
intrinsically magnetized, with the magnetization given by
$\vecM_V (\vecr) = (\gamma_n\hbar) \vecsigma(\vecr)/2 = g_n\mu_N \vecsigma(\vecr)/2$, where
$\gamma_n = g_n\mu_N\hbar^{-1}$ is the gyromagnetic ratio of neutron
and $g_n= −3.826$, $\hbar\vecsigma/2$ is the spin density which can
be estimated for the $P$-wave vortex as \cite{1982PhRvD..25..967S} 
\bea\label{eq:Pairing_spin_density}
\sigma(r) = \frac{\nu_n \Delta_n^2}{3} \ln
\left(\frac{\Lambda}{T_c}\right) g(r) [f_1(r)-f_2(r)], 
\eea
where $\Lambda$ is the BCS cut-off, $\nu_n$ - the neutron density of states at
the Fermi surface. The magnitude of the vortex magnetization that
follows from Eq.~\eqref{eq:Pairing_spin_density} is estimated as
\cite{1982PhRvD..25..967S}
\bea
\vert M_V(^3P_2)\vert = \frac{g_n\mu_N}{2} n_n\left(\frac{\Delta}{\ep_{Fn}}\right)^2
\simeq 10^{11} ~{\textrm G}.
\eea
Ambient electrons which coexist with the $P$-wave superfluid because
of approximate  $\beta$-equilibrium among neutrons, protons and
electrons, will scatter off the magnetized vortices via the QED
interaction  term $-e\vecgamma\cdot \vecA (\vecr)$, where the vector
potential associated with the vortex is given by $\vecA(\vecr) =
A(r)\hat\phi$
\bea 
A(r) = \frac{1}{r} \int_0^r \vert M_V(r')\vert r'dr'. 
\eea 
The relaxation time for the electron-vortex scattering is given by
\bea\label{eq:P_wave_relax}
\tau_{eV}[^3P_2] \simeq \frac{7.91 \times 10^{8}}{\Omega_s}
\left(\frac{k_{Fn}}{\textrm{fm}}\right)
\left(\frac{\textrm{MeV}}{\Delta_n}\right)
\left(\frac{n_e}{n_n}\right)^{2/3}.
\eea
An important feature of this relaxation time is its near independence
of the temperature (a weak temperature dependence arises because of
the temperature dependence of the gap). Therefore, it provides an
asymptotic lower limit on the scattering rate at low temperatures
($T\ll \Delta_n$) where the relaxation time $\tau_{eV}[^1S_0]$ given
by Eq.~\eqref{eq:S_wave_relax} is exponentially
suppressed. Numerically $\tau_{eV}[^3P_2]$ is of the order of tens of
days for the period of the Vela pulsar and varies weakly with the
density within the core region where $P$-wave superfluid resides.

So far, for simplicity, we neglected the proton component of the core
of a neutron star. However, as we describe below, the proton fluid can
dramatically modify the mutual friction in the core of the
star. Consider first a normal (non-superconducting) fluid of
protons. At high densities the proton energies can indeed become large
enough so that the $^1S_0$-wave interaction becomes negative and
pairing vanishes. As described in Sec.~\ref{sec:Pairing_Strong_B},
strong magnetic fields $B\ge H_{c2}\simeq 10^{15}$ G unpair the proton
fluid. Non-superconducting protons will couple to electrons on short
plasma time-scale, the relevant scale being set by the plasma
frequency. Therefore, protons will compete with electrons in providing
the most efficient interaction with the neutron vortices and,
eventually, the coupling between the charged plasma component and the
neutron superfluid. The key advantage of protons over electrons is
that they couple to neutrons via the strong nuclear force, instead of
much weaker electromagnetism. The relaxation time for the proton
scattering off the quasiparticles in the cores of $S$-wave neutron
vortices is given by~\cite{1998PhRvD..58b1301S}
\bea\label{eq:np_relax}
 \tau_{pV}[^1S_0] = \frac{0.71}{\Omega_s} 
\frac
{m_n^*m_p^*}{m_n\mu_{pn}^*}\left(\frac{\ep_{Fp}}{\ep_{Fn}}\right)^2
\frac{\ep_{1/2}^0}{T} \exp\left(\frac{\ep^0_{1/2}}{T}\right)
\frac{\xi_n^2}{\langle \sigma_{np}\rangle},
\eea
where $\mu_{pn}^* = m_p^*m_n^*/(m_n^*+m_p^*)$ is the reduced mass of
the neutron-proton system (entering the relation between the cross-section
and the scattering amplitude squared), $\ep^0_{1/2}$ is the lowest
eigenstate of vortex core excitations,
Eq.~\eqref{eq:vortex_eigenstate}, and $\langle \sigma_{np}\rangle$ can
be viewed as an average neutron-proton cross-section (for a more
precise definition see Eq. (20) of Ref.~\cite{1998PhRvD..58b1301S}).
The bare neutron mass $m_n$ stems from the quantum of circulation
$\kappa=\pi\hbar /m_n$ defining the number of vortices in terms of spin
frequency $\Omega_s$. Numerical evaluation of Eq.~\eqref{eq:np_relax} shows
that in the relevant range of temperatures $T\simeq 10^{7}$-$10^{8}$~K
the relaxation time is of the order $\tau_{pV}[^1S_0] \simeq 10^{-2}$
sec for the period of the Vela pulsar, which is much shorter than the
time-scales for electromagnetic coupling of electrons given by
Eqs.~\eqref{eq:S_wave_relax} and \eqref{eq:P_wave_relax}. Only at
temperatures of the order of several $10^6$~K the process
\eqref{eq:P_wave_relax} takes over; however such low temperatures are
unlikely to be achieved in neutron star cores. One potentially
important consequence of the shortness of the relaxation time
\eqref{eq:np_relax} is that magnetars superfluid cores will couple
to the remaining stellar plasma on short dynamical time-scales once
 proton superconductivity is suppressed by the unpairing mechanisms
discussed in Sec.~\ref{sec:Pairing_Strong_B}. This will limit the
possibilities of explaining glitches and long-timescale variability
in terms of superfluid dynamics of magnetar
cores~\cite{2016A&A...587L...2S}.

Next let us turn to the case where protons are superconducting. 
A fundamentally new aspect in this case is the {\it entrainment} of the
proton condensate by the neutron condensate which leads to
 magnetization of a neutron vortex by protonic entrainment 
currents~\cite{1981Vardanyan,1984ApJ...282..533A}. The
effective flux of the neutron vortex is given by 
\bea\label{eq:effective_flux}
\phi^* = k\phi_0, \quad k = \frac{m_p^*}{m_p},
\eea
where $\phi_0 = \pi\hbar c/e$ is the quantum of flux. Numerically, the
magnitude of the field in this case is by four orders larger than due
to the spontaneous magnetization~\cite{1980PhRvD..21.1494M}, therefore
much shorter relaxation times are expected.  The calculation of the
electron relaxation on a neutron vortex discussed in the case of
$P$-wave superfluid can be repeated in the case of the magnetization
by entrainment currents~\cite{1984ApJ...282..533A}. It is convenient
to define the zero-radius scattering rate as
\bea
\tau_{0}^{-1} &=& \frac{ 2c n_v  }{k_{eF}}\left(\frac{\pi^2\phi_*^2}{4\phi_0^2}\right),
\eea
where the term separated in the bracket is an approximation to the
exact Aharonov-Bohm scattering result where $\sin^2
(\pi/2)(\phi_*/\phi_0)$ appears instead. 
The full finite-range scattering rate is then given
by~\cite{1984ApJ...282..533A}
\bea
\tau^{-1}_{e\phi} = \frac{3\pi}{32} \left(\frac{\ep_{Fe}}{m_pc^2}\right)
\frac{\tau_{0}^{-1} }{k_{eF}\lambda},
\eea
where $\lambda$ is the penetration depth.   As was the case with the
relaxation time \eqref{eq:P_wave_relax} the scattering rate from
magnetized neutron vortices is temperature independent in the first
approximation. Numerically, the relaxation time is within the range of
seconds and is about four order of magnitude shorter than the one
given by \eqref{eq:P_wave_relax}, as a direct consequence 
of the induced field being  larger by the same amount.

As discussed in Sec.~\ref{sec:Pairing_Strong_B}, the magnetic field of
a neutron star will penetrate the superconducting proton fluid by
either forming quantized vortices (in the case where the proton
superconductor is type-II) or domains of unpaired proton matter (in
the case where it is type-I). Consider first a type-II
superconductor. For fields of the order of $10^{12}$~G and typical
rotation frequencies of neutron stars ($\Omega\simeq 100$ Hz), the
number of proton vortices (or flux tubes) per area of a single neutron
vortex (assuming for the sake of argument colinear neutron and proton
vortex lines) is of the order of $10^{13}$. Therefore, one may
anticipate that proton vortices might strongly affect neutron vortex
dynamics. There exist several scenarios on how neutron and proton
vortex systems interact and we review them in turn.

One possible scenario assumes that the proton vortex network continues
into the crust via magnetic field lines and therefore is frozen into
the crustal electron-ion plasma. Neutron vortices might then get
pinning on or between these vortices because of the long-range
hydrodynamical interaction between them on characteristic scales of
the order of $\lambda$. Microscopically, crossing the vortices may
lead to additional pinning force on the scale of $\xi$ where the
condensate is quenched~\cite{1985Ap&SS.115...43M,1989ASIC..262..457S},
but the long-range $\sim \lambda$ the hydrodynamical force is the dominant
component.  The pinning of neutron vortices to proton flux tubes may
thus substantially affect the dynamics of neutron vorticity and to
some extent can be viewed as mutual friction. The magnitude of the
pinning force strongly depends on the relative orientation of vortices
and flux tubes, which does not permit to draw model independent
conclusions on the relative importance of the pinning force. In some
models there are flux tubes associated with the different components
of the magnetic fields (poloidal, toroidal, etc), therefore the
geometry of the flux tube network itself is a complicated problem.
These type of models will be discussed in more detail in
Sec.~\ref{sec:pinning} below.

In the vortex cluster model~\cite{1991SvPhU..34..555S} a neutron vortex
carries a cluster of proton vortices colinear with the neutron vortex,
which are arranged within the region where the entrainment induced
field exceeds the lower critical field of the proton superconductor
$H_{c1}$. Such a cluster may substantially enhance the scattering rate
estimate given in Ref.~\cite{1984ApJ...282..533A}. Larger scattering
rate and large forces on the neutron vortex from the electron fluid can
lead (counter-intuitively) to longer relaxation times for the neutron
superfluid than predicted in Ref.~\cite{1984ApJ...282..533A}; as a
consequence post-glich relaxation time-scales appear to be compatible
with the vortex cluster model~\cite{1995ApJ...447..305S}.

Understanding of mutual friction in the case of type-I superconducting
protons is difficult because of a lack of model-independent predictions
for the domain structure and size of type-I superconductor. A
tractable case is where a neutron vortex carries a colinear normal
proton domain; in this case it can be shown that the neutron vortex
motion induces an electric current within the domain which leads in turn
to Ohmic dissipation of electron 
current~\cite{2005PhRvD..71h3003S}. The dimensionless drag to lift
ratio for this process was found to be of the order of $10^{-4}$,
which makes precession in neutron stars compatible with the type-I
superconductivity of protons~\cite{2003PhRvL..91j1101L}.

We now turn to the discussion of the mutual friction in the $S$-wave
superfluid within the neutron star's crust. Here the lattice of crustal
nuclei is the physically distinct new component which was absent in
our discussion of the core physics. Because in the crust the protons
are confined to the nuclei and the superfluid is $S$-wave, the only
process that can be carried over from the discussion above is the one
give by Eq.~\eqref{eq:S_wave_relax}. However, according to our
current understanding it is not the dominant process of mutual friction in
the crust. 

The stationary state of a neutron vortex might require its pinning on
a nucleus or in the space between nuclei. As we shall see in section
\ref{sec:pinning} there are substantial differences in the estimates
of the pinning force in the crust, however the most advanced
treatments of pinning based on the density functional theory of
superfluid matter indicate that the pinning occurs between the
nuclei~\cite{2016PhRvL.117w2701W}. The sign of the pinning force
makes, however, little difference; if the pinning of vortices is
strong, then these can respond to the changes in the rotation rate of
the crust via thermally activate creep~(see
Ref.~\cite{2014ApJ...789..141L} and references therein). Vortex
creep theory postulates a form of the radial velocity of a vortex
which depends exponentially on the ratio of the pinning pontential to
the temperature. If pinning is strong the associated drag-to-lift
ratio could be large and in the range that can account for long
time-scale relaxations of glitches~\cite{2014ApJ...789..141L}.

The strong pinning regime may not arise when the vortex lattice is
oriented randomly with respect to the basis vectors of the nuclear
lattice.  A neutron vortex moving in the crust will couple to the
phonon modes of the nuclear lattice~\cite{1990MNRAS.243..257J}.  The
one-phonon processes lead to a weak coupling of the superfluid to the
crust with $\eta \simeq 10$ g cm$^{-1}$ s$^{-1}$ which implies small
dimensionless drag-to-lift ratio $\eta /\rho_n\kappa\ll 1$.  

The interaction of a neutron vortex moving relative to the nuclei in
the crust will generate oscillation modes (Kelvin modes or kelvons) on
the vortex and will thus disspate the kinetic energy of vortex motion
into the energy of oscillations~\cite{1992ApJ...387..276E}.  This
dissipation can be viewed as mutual friction, because energy 
and  momentum is transferred between the superfluid and the
crust. Ref.~\cite{1992ApJ...387..276E} expresses the drag-to-lift
ratio in terms of a dissipation angle
$\eta/\rho_n\kappa = \tan \theta_d$ and finds an upper limit on this
angle $\theta_d \le 0.7$. This implies that  throughout the crust this 
dissipation mechanism leads to dissipation angles and spin-up
time-scales which are close to the maximal one
$t_{\rm s,max} \approx (2\Omega)^{-1}.$  It has been argued that the
randomness of nuclear potentials may suppress the kelvon-excitation
mechanism~\cite{1992MNRAS.257..501J} with two-phonon processes being
important in a certain range of parameters.

Because the relative orientation of the circulation vector of the vortex
lattice and the crustal lattice basis vectors may be random or
dependent on the history of solidification of the crust and nucleation
of the superfluid phase, it remains an open question whether the
pinned regime is realized in the crust of a neutron stars. Numerical
simulations~\cite{2016PhRvL.117w2701W} and/or astrophysical
constraints coming from glitch observations may eventually distinguish
between the various models.

\section{Superfluid hydrodynamics}
\label{sec:superfluid_hydro}

Let us now discuss how to model a superfluid star on the macroscopic,
hydrodynamical scale. In order to do hydrodynamics it is necessary to
average over length-scales that are large enough for the constituents
to be considered `fluids'. In the case of superfluids this means
course graining not only over length-scales much larger than the
mean free path of the particles, but also over length scales much 
larger than the characteristic scale of vortices or flux tubes. 

A superfluid forms a macroscopic coherent state, therefore it can be
described by a {\it macroscopic} wave function
$\psi(\vecr)=|\psi(\vecr)|\exp[{i\chi(\vecr)}]$, which implies that
the number density of the superfluid is given by
$\vert\psi(\vecr)\vert^2$ and the momentum is just the gradient of the
phase $\vecp = \hbar \vecnabla \psi(\vecr)$. This implies immediately
that
\bea\label{eq:quantization}
\vecnabla\times\vecp =\vecnabla\times\vecnabla\chi(\vecr)=0,
\eea
i.e., the superfluid is irrotational. However, rotating superfluids
support rotation by forming quantized vortices, above certain critical
angular velocity $\Omega_{c1}$.  Indeed, the minimization of the free
energy of the fluid in the rotating
frame~\cite{khalatnikov1989introduction}, i.e.,
\bea
F_r=F-\Omegavec \cdot \vecJ, 
\eea
where $\vecJ$ the angular momentum of the fluid and $F$ is its free
energy in the laboratory frame, leads to a solution predicting rigid
rotation, which is supported by the condensate through the formation of an
array of superfluid vortices. The circulation of a single vortex is
quantised as
\bea
\oint\vecp\cdot d\ \vecl =m_n \ n\  \kappa,\quad n = 1,2,\dots
\eea
where $\kappa=\pi\hbar/m_n\approx 2\times 10^{-3}$ cm$^2$ s$^{-1}$ is
the quantum of circulation and $m_n$ is the mass of the neutron.

In neutron stars the hydrodynamical description of neutron fluid
requires thus averaging over length-scales larger than the
inter-vortex separation, in order to define the rotation rate by
averaging over many vortices.  This means that in general a fluid
element must be small compared to the radius of the star
($R \simeq 10$ km), but large compared to the inter-vortex separation
$d_n\simeq 10^{-3} \left(P\mbox{/10 ms}\right)^{1/2}$~cm, with
$P = 2\pi/\Omega$ being the spin period of the star.

In a realistic neutron star one has to account for multiple
superfluid/superconducting fluids,  and a minimal model, such as we
now present, must account at least for a neutron-proton
conglomerate with the background of electrons. Consider first the
simpler Newtonian
case~\cite{1981Vardanyan,1991AnPhy.205..110M,1991ApJ...380..515M,1995ApJ...447..305S,1998MNRAS.296..903M,2011MNRAS.410..805G,2015MNRAS.453..671G,2017MNRAS.465.3416P,
2017MNRAS.464..721A}.  In
the case of interest to us, the indices $\x, \ \y$ label either
protons ($\p$) or neutrons ($\n$) and electrons are assumed to form a
charge neutralizing background moving with proton fluid on timescales
much shorter than those of interest here, such as dynamical and
oscillation timescales of milliseconds, or glitch timescales of
minutes or more \cite{1991ApJ...380..515M}.

We will not list the complete set of magneto-hydrodynamics equations
here and will concentrate on the ingredients that are needed in the
modeling the dynamics of a rotating and magnetized neutron star
following Ref.~\cite{2006CQGra..23.5505A}.
Corrections due to coupling of the superfluid
to its excitations are ignored. These can be accounted for by
including an additional fluid of excitations to the analysis, which
also allows to recover the standard hydrodynamical equations used for
superfluid helium ~\cite{2004PhRvD..69d3001P, 2012PhRvD..86f3002H}.
Each constituent $\x$
conserves its mass density $\rho_\x$ (in the following a summation
over latin indices is assumed)
\be
\partial_t\rho_\x+\nabla_i(\rho_\x v_\x^i)=0,
\ee
and Euler equations for their velocities $v_\x^i$ are given by
\be
(\partial_t+v_\x^j\nabla_j)(v_i^\x+\varepsilon_\x
w_i^{\y\x})+\nabla_i(\tilde{\mu}_\x+\Phi)+\varepsilon_\x
w_{\y\x}^j\nabla_i v^\x_j=(f_i^{\x, \mf}+f_i^{\x, \pin}+f_i^{\x, 
\mg })/\rho_\x,
\label{euler}
\ee
where $w_i^{\y\x}=v_i^\y-v_i^\x$, $\tilde{\mu}_\x=\mu_\x/m_\x$ is the
chemical potential per unit mass.
The key new aspect of these treatments is the
{\it entrainment} effect which causes the momentum and
the velocities of the components to not be parallel, but rather related by 
\bea
p_i^\x=m_\x (v_i^\x+\varepsilon_\x
w_i^{\y\x}).
\eea
This observation was first made by Andreev and Bashkin in the context of
charge neutral $^3$He-$^4$He mixtures~\cite{1976JETP...42..164A}.
Here $\varepsilon_\x = 1-m^*_\x/m_\x$
is the entrainment coefficient which accounts for the non-dissipative
coupling between the components. (It is related to quantity $k$
defined in \eqref{eq:effective_flux} by 
$\varepsilon_\x = 1-k$).  Note that the microphysical calculations
predict effective mass $m^*_\x\lesssim m_\x$ in the core, in which
case the entrainment coefficient is positive. The opposite relation
holds in the curst of a neutron stars, i.e.  $m^*_\x \gg m_\x$, due to
the band structure of the nuclear lattice and Bragg scattering
\cite{2012PhRvC..85c5801C}.

The gravitational potential $\Phi$ in Eq.~\eqref{euler} obeys the
Poisson equation
\be
\nabla^2 \Phi = 4\pi G \sum_\x \rho_\x.
\ee
On the right-hand side of Eq.~\eqref{euler} the forces are decomposed
as follows. The first term $f^{\x, \pin}_i$ is the force due to pinned
vortices, $f_i^{\x, \mf}$ is the mutual friction force mediated by
free vortices, and $f_i^{\x, \mg}$ is the force due to the magnetic
field, which as we shall see depends strongly on whether the protons
are superconducting or not.

We will discuss the contributions from the mutual friction and
magnetic forces in detail below in this section, whereas the pinning
force is discussed in Sec.~\ref{sec:pinning}.  For laminar flows and
straight vortices, the mutual friction force has the standard
Hall-Vienen-Bekarevich-Khalatnikov form
\cite{1956RSPSA.238..215H,khalatnikov1989introduction}
\be
f_i^{\x, \mf}=\kappa n_\vv \rho_\n \mathcal{B}^{'}\epsilon_{ijk}\hat{\Omega}^i_\n w^k_{\x\y}+\kappa n_\vv\rho_\n\mathcal{B}\epsilon_{ijk}\hat{\Omega}^j_\n\epsilon^{klm}\hat{\Omega}^\n_l w_m^{\x\y},
\label{MF}
\ee
where $\Omega_\n^j$ is the angular velocity of the neutrons (a hat
represents a unit vector) and $\mathcal{B}$ and $\mathcal{B}'$ are the
mutual friction coefficients.  The neutron vortex density per unit
area is $n_\vv$, and is linked to the rotation rate (at a cylindrical
radius $\varpi$) by the relations
\be
\kappa n_\vv (\varpi) = 2\tilde \Omega + \varpi \frac{\partial}{\partial \varpi}
\tilde \Omega, 
\qquad \tilde\Omega \equiv \Omega_\n + \varepsilon_\n (\Omega_\p -
  \Omega_\n),
\label{MutF}
\ee
obtained by imposing that the circulation derived by integrating over
a contour the smoothed average momentum is the sum of the quantised
circulations of the $\mathcal{N} (\varpi)$ vortices enclosed, i.e.
\be
\oint \epsilon^{ijk}\nabla_j p_k^\n=2 \pi \int_0^\varpi 
m_n\tilde \Omega  r  dr
=\mathcal{N}(\varpi) m_\n \kappa,
\ee 
and we assume here and below singly quantized vortices.
The parameters $\mathcal{B}$ and $\mathcal{B}^{'}$ depend on the
microphysical processes giving rise to the mutual friction, as
described in Sec.~\ref{sec:mutial_friction}, and can be expressed in
terms of a dimensionless drag-to-lift ratio parameter $\mathcal{R}$
(see also Sec.~\ref{sec:mutial_friction}) related to the dimensionfull
drag parameter $\eta $ [g cm$^{-1}$ s$^{-1}$] as
\be
{\mathcal R}=\frac{\eta}{\kappa \rho_\n},
\ee
according to 
\be
\mathcal{B}=\frac{\R}{1+\R^2},
\qquad 
\mathcal{B}^{'}=\frac{\R^2}{1+\R^2}.
\ee
To connect to the discussion of the relaxation times computed in
Sec.~\ref{sec:mutial_friction}, we now express $\eta$, or equivalently
${\mathcal B}$ and ${\mathcal B'}$ in terms of these microscopic
time-scales.  We distinguish two cases of non-relativistic and  
and ultra-relativistic unpaired excitations, which we assume 
to be protons ($\p$) or electrons ($\e$).
The force exerted by non-superconducting quasiparticles
per single vortex is given in general by~\cite{1989ApJ...342..951B}
\be \label{eq:friction_force} \vecf_d =
\frac{2}{\tau n_\vv} \int f (\vecp, \vecv_\vv) \vecp
\frac{d^3p}{(2\pi\hbar)^3} = -\eta \vecv_\vv, 
\ee
where $\vecv_\vv$ is the velocity of the vortex in a frame co-moving with the
normal component and $f (\vecp, \vecv_\vv)$ is the non-equilibrium distribution
function, which we expand assuming small perturbation about the
equilibrium distribution function $f_0$, that is,
$                                                                                                           
f (\vecp, \vecv_\vv) =                                                                                        
 f_0 (\vecp)+ (\partial f_0/ \partial  \epsilon)  (\vecp\cdot \vecv_\vv).                                     
$
In the low-temperature limit
$\partial f_0/\partial \epsilon\simeq -\delta(\epsilon-\epsilon_{F})
$,
where $\ep_F$ is the corresponding Fermi energy.  After integration
one finds
\bea \eta_{\p} =m_{\p}^* \frac{ n_{\p}}{\tau_{\p} n_\vv}, \qquad \eta_{\e} =
\frac{\hbar k_\e}{c} \frac{n_\e}{\tau_{\e} n_\vv} ,\eea
where $n_{\p,\e}$ are the proton/electron number densities. Note that
if both electron and proton quasiparticles are present then the
contributions from $\eta_\e$ and $\eta_\p$ need to be summed, just as
in the case of the ordinary transport coefficients.

The parameters $\mathcal{B}$ or
${\mathcal R}$ can be extracted from the timescales obtained in
Sec.~\ref{sec:mutial_friction}, using
\bea
\tau_{\rm mf} = \frac{1}{2\Omega_s(0) {\mathcal B}}\;\;,
\eea
where $\Omega_s(0)$ is the spin frequency of the superfluid at $t=0$ .

From the equations in (\ref{euler}) we can also see that in the case of two constant density rigidly rotating fluids, with
moments of inertia $I_\p$ and $I_\n$ and frequencies $\Omega_\n$ and $\Omega_\p$, and
neglecting for the sake of the argument the external torques and entrainment, an
initial difference in rotation rate $\Delta\Omega=\Omega_\p-\Omega_\n$
will be erased by mutual friction according to
$\Delta\Omega(t)=\Delta\Omega(0)\exp({-t/\tau_{\rm su}})$
~\cite{1989ApJ...342..951B, 2006MNRAS.368..162A}, with
\bea
\tau_{\rm su} \approx \left(\frac{I_\p}{I_\n+I_\p}\right) \frac{1}{2\Omega_\n (0) {\mathcal B}}\;\;,
\eea
where $\Omega_\n(0)$ is the spin frequency of the neutron fluid at $t=0$.

The expression for the mutual friction in Eq.~(\ref{MutF}) is appropriate
for straight vortices in a triangular array which corresponds to the
minimum of the free energy of a rotating superfluid. However, vortices 
are likely to bend due to their finite rigidity and this effect
can easily be included in the expression for the mutual friction, see
Ref.~\cite{khalatnikov1989introduction} for a discussion in single
component fluids and its extension to multi-component fluids in
Refs.~\cite{1991ApJ...380..515M,1995ApJ...447..305S,2007MNRAS.381..747A}.
Furthermore, it is well known from laboratory experiments with
superfluid $^4$He 
that a counterflow along the vortex axis can trigger the
Glaberson-Donnelly instability
\cite{1974PhRvL..33.1197G,1991qvhi.book.....D} and destabilise the
vortex lattice, creating a turbulent tangle.  In the case of an
isotropic tangle a phenomenological form for the mutual friction, due
to Ref.~\cite{1949Phy....15..285G}, is
\be
f_i^{GM}=\frac{8\pi^2\rho_\n}{3\kappa}\left(\frac{\xi_1}{\xi_2}\right)^2\mathcal{B}^3 w^2_{\p\n} w_i^{\p\n},
\label{GM}
\ee
where the phenomenological parameters are set to $\xi_1\approx 0.3$
and $\xi_2\approx 1$. In neutron star interiors the presence of large
relative flows between the `normal' and superfluid components and
large Reynolds numbers, of the order of Re$\geq 10^7$ are likely to
lead to superfluid turbulence and the presence of a vortex tangle.
According to
Refs. \cite{2006ApJ...651.1079P,2007ApJ...662L..99M,2007MNRAS.381..747A}
a polarized tangle is expected in a rotating pulsar.

Let us shift our attention to the magnetic force. The equations of
magneto-hydrodynamics for a superfluid and superconducting neutron
star have been initially considered in
Refs.~\cite{1981Vardanyan,1991AnPhy.205..110M,1991ApJ...380..515M,1995ApJ...447..305S,1998MNRAS.296..903M}. More
recently detailed studies were carried out
Refs.~\cite{2011MNRAS.410..805G,2015MNRAS.453..671G,2017MNRAS.465.3416P}
which to various degree also include discussion of the evolution of
the magnetic field in a superconducting neutron star.  For the current
discussion let us restrict out attention to the simplified case of a
two component neutron star, in which the electrons are assumed to move
with the protons. In this case one finds
\beq
f^i_{\p, \mg}&=&\frac{1}{4\pi} \left[ B^j \nabla_j (H_{c1}\hat{B}^i)-B\nabla^i H_{c1}\right] -\frac{\rho_\p}{4\pi}\nabla^i\left(B\frac{\partial H_{c1}}{\partial \rho_\p}\right),\\
f^i_{\n, \mg}&=&\frac{1}{4\pi} \left[\mathcal{W}_\n^j\nabla_j (H_{v\n}\hat{\mathcal{W}}_\n^i)-\mathcal{W}_\n\nabla^i H_{v\n}\right]-\frac{\rho_\n}{4\pi}\nabla^i\left(B\frac{\partial H_{c1}}{\partial \rho_\n}\right),
\eeq
where a hat indicates a unit vector and we have defined
\be
\mathcal{W}^i_\p=\epsilon^{ijk}\nabla_j(v^\p_k+\varepsilon_\p w_k^{\n\p})+a_p B^i= n_{\mathrm{vp}} k^i_\p
\ee
with $k^i=\kappa \hat{k}^i$ pointing along the local vortex direction,
$a_\p=e/mc\simeq 9.6\times 10^3$ G$^{-1}$ s$^{-1}$ and
$n_{\mathrm{vp}} $ the surface density of proton vortices.
The total magnetic induction $B^i$ is the sum of three terms
\be
B^i=B_\p^i+B_\n^i+b_L^i,
\ee
where $B_\p^i$ is the contribution due to the proton vortices,
$B_\n^i$ is the contribution due to the neutron vortices and $b_L^i$
is the London filed. The modulus of the induction is
$B=\sqrt{B_iB^i}$. Generally 
$|B_\p| \gg |B_\n| \approx |b_L|$, i.e.,  the strengths of both the
average field due to neutron vortices and the London field is
completely negligible compared to neutron star interior magnetic
fields ($|B_\n| \approx |b_L|\approx 10^{-2}$ G). The term
$H_{v\n}=4\pi a_\p {\mathcal{\varepsilon}_{v\n}}/\kappa \approx 10
\times H_{c1}$
plays the role of an effective magnetic field
\cite{2011MNRAS.410..805G} and depends on the energy per unit length
of a neutron vortex
\be
\varepsilon_{v\n}\approx \frac{\kappa^2}{4\pi}\frac{\rho_\n}{1-\varepsilon_\n}\log\left(\frac{l_\vv}{\xi_\n}\right),
\ee
where $\xi_\n$ is the coherence length of the vortex and $l_\vv$ the inter-vortex separation, and we can approximate $\log (l_\vv/\xi_\n)\approx 20-1/2\log(\Omega/100 \mbox{ rad/s})$ \cite{2007MNRAS.381..747A}.
To study the coupled evolution of the fluid and magnetic field these
equations need to be coupled to the induction equations for the
magnetic field:
\be
\partial_t B^i = \epsilon^{ijk}\epsilon_{klm} \nabla_j (v_\e^l B^m)-\frac{1}{a_\p}\epsilon^{ijk}\nabla_j f_k^\e,
\ee
where, under the assumption that $n_{\mathrm{vp}}\gg n_\vv$ one finds
\be
f_\e^i\approx \frac{1}{4\pi\rho_\p}\frac{\R_\p}{1+\R^2_\p}\left[\R_\p{B^j\nabla_j(H_{c1}\hat{B}^i)-B\nabla^i H_{c1}}-\epsilon^{ijk} B_j\nabla_k H_{c1}\epsilon^{ijk}\hat{B}_j B^l \nabla_l \hat{B}_k \right],
\ee
where $\R_\p$ is the drag parameter describing the scattering of electrons
on proton vortices.

One may also expect an additional contribution to the mutual
friction, due to the flux-tube and neutron vortex interaction, of the form
\beq
f^i_{\p\n, \mf}&=&\rho_\n \kappa n_\vv \frac{\mathcal{C}_\vv}{1+\R_\p^2}[\R_\p f_{*}^i + \epsilon^{ijk} \hat{\mathcal{W}}^\p_j f_{*k}],\\
f_{*}^i&\approx&\frac{1}{4\pi a_\p \rho_\p} [\hat{B}^j\nabla_j (H_{c1}\hat{B}^i)-\nabla^i H_{c1}],
\eeq
where $\mathcal{C}_\vv$ is a phenomenological coefficient that
parameterises the strength of the resistive interaction between the
proton and neutron vortex arrays, which may also drive the evolution
of the magnetic field \cite{1998ApJ...492..267R}.

The magnetic field configuration of superfluid and superconducting
neutron stars has been analysed in detail in recent years both by
studying equilibrium models \cite{2013PhRvL.110g1101L,2013MNRAS.431.2986H,
  2014MNRAS.437..424L} and, more recently, by studying the evolution
of the coupled core and crust magnetic fields
\cite{2016MNRAS.456.4461E}. In general superfluidity and
superconductivity have a strong impact on the timescales for the
evolution of the core magnetic field \cite{2017arXiv170402016P}, and
for strongly magnetised neutron stars ($B\geq 10^{14}$ G) could lead
to the expulsion of the toroidal field from the core, and significant
rearrangement of the crustal magnetic field on timescales comparable
to, or shorter than, the age of the star.

\subsection{Relativistic fluids}
\label{sec:relativity}

What has been presented up to now is the Newtonian framework for
describing superfluid and superconducting neutron stars. One can
develop a similar framework in general
relativity~\cite{1992AnPhy.219..243C,1994RvMaP...6..277C,1998NuPhB.531..478C},
for a review see Ref.~\cite{2007LRR....10....1A}.  First define the
number density four-currents of each component
\bea
n^\mu_\x=n_\x u^\mu_\x,
\eea
with normalization $u^\x_\mu u^\mu_\x=-1$ and Greek letters
representing four dimensional space-time indices; summation is
implicit over repeated Greek indices.  A master function $\Lambda$ is then
defined which is a function of the scalars of the system, in
particular the number densities $n_\x$ and
$n_{\x\y}^2=n_{\y\x}^2=-g_{\mu\nu} n_\x^{\mu} n_\y^\nu$, where
$g_{\mu\nu}$ is the metric \cite{1994RvMaP...6..277C}.  In the case of
co-moving fluids $\Lambda$ is (up to the sign) simply the local
thermodynamical energy density. In the general case $\Lambda$ includes
relative flows of the fluids by definition.

Having at our disposal the master function we can proceed to define
the conjugate momenta in the standard fashion of the Lagrangian
theory~\cite{1992AnPhy.219..243C,1994RvMaP...6..277C,1998NuPhB.531..478C}
\beq
\pi_\mu^\x&=&g_{\mu\nu} (\mathcal{A}^\x n_\x^\nu+\mathcal{A}^{\x\y} n_\y^{\nu}),\\
\mathcal{A}^\x&=&-2\frac{\partial\Lambda}{\partial n_\x^2},\\
\mathcal{A}^{\x\y}&=&\mathcal{A}^{\y\x}=-\frac{\partial\Lambda}{\partial n_{\x\y}^2},
\eeq
where the effect of entrainment is encoded in the coefficients
$\mathcal{A}^{\x\y}$. The stress-energy tensor is defined as 
\be
T^\mu_\nu=\Psi \delta^\mu_\nu-\sum_\x n_\x^\mu\pi^\x_\nu,
\ee
with the generalised pressure $\Psi$ defined as
\be
\Psi=\Lambda-\sum_\x n_\x^\mu \pi_\mu^\x.
\ee
The equations of motion for the fluid can then be written as a set of
Euler equations 
\be
\sum_\x n_\x^\mu\nabla_{[\mu}\pi_{\nu]}=0,
\ee
to be solved together with the Einstein's equations of general
relativity and conservation equations for the individual four-currents
of the components
\be
\nabla_\mu n^\mu_\x=0.
\ee
Note that the solution to the equations above automatically satisfies
the energy-momentum conservation $\nabla_\mu T^{\mu}_{\nu}=0$. 

Relativistic superfluid hydrodynamics of the type described above was
formulated initially in
Refs.~\cite{1992AnPhy.219..243C,1994RvMaP...6..277C,1998NuPhB.531..478C}.
The corresponding equations where adapted to differentially rotating
relativistic superfluid neutron stars in
Ref.~\cite{1998MNRAS.297.1189L} in the case of cold equations of
state. Equilibrium configurations of two-fluid (superfluid and normal
component featuring) neutron stars were constructed in
Refs.~\cite{2005PhRvD..71d3005P,2016PhRvD..93h3004S}.  Accounting for
heat transport, dissipation and in particular vortex-mediated mutual
friction is more challenging in general relativity than in Newtonian
physics \cite{2013MNRAS.428.1518G}, as standard approaches by
Refs.~\cite{1940PhRv...58..919E,1959flme.book.....L} lead to
causality and stability problems. While
Refs.~\cite{1979AnPhy.118..341I,1991RSPSA.433...45C}
resolve some of these issues, and progress has been made making
maximal use of the variational approach \cite{2015CQGra..32g5008A} the
general relativistic formulation is, however, not complete.  Some
recent advances in the problem of mutual friction and vortex motion in
a relativistic framework can be found in
Refs.~\cite{2016CQGra..33x5010A,2016PhRvD..93f4033G}.

\section{Pinning effects}
\label{sec:pinning}

In the previous discussion we have considered mainly vortices that are
free to move with respect to the fluid components and experience a
standard drag force, linear in the difference in velocity between said
component and the vortices themselves. (A brief discussion of pinning
in the crust was given at the end of Sec.~\ref{sec:mutial_friction} to
complete the discussion of microphysics of mutual friction.)  However
the interaction between vortices and ions in the crust, or flux-tubes
in the core, can be strong enough to balance the Magnus force, and
`pin' the vortices, preventing them from moving, similarly to static
friction. We now turn to the detailed discussion of the pinning
effects in the superfluid core and in the crust of the neutron star.

\subsection{Vortex-flux tube pinning and interactions}
\label{sec:fluxtubes}

In the outer core of the neutron star, where neutrons are superfluid
and protons are expected to form a type II superconductor, the
magnetic field is confined to flux-tubes with flux quantum
$\phi_0=\pi\hbar c/e \approx 2 \times 10^{-7}$~G~cm$^{2}$. The neutron
vortices thus co-exist with an array of far more numerous flux tubes,
with average spacing
\be
l_{\phi}=\frac{B}{\phi_0}\approx 4 \times 10^{3} 
\left(\frac{B}{10^{12} \mbox{ G}}\right)^{-1/2} \mbox{fm}.
\ee
Proton flux tubes are also less rigid than neutron vortices. Indeed
the tension of a neutron vortex  is given by 
\be
T_v=\frac{\rho_\n\kappa^2}{4\pi} \ln \left(\frac{l_v}{\xi_n}\right)\approx 10^9
\left(\frac{\rho_\n}{2\times 10^{14} \mbox{ g cm$^{-3}$}}\right) 
\mbox{erg cm}^{-1}
\label{Tension}
\ee
with the typical inter-vortex separation 
$l_v\approx  10^{-3} \left(\mbox{P/10 ms}\right)^{1/2}$ cm.
The flux-tube tension is given by \cite{1986PhRvD..33.2084H}
\be
T_\Phi=\left(\frac{\phi_0}{4\pi \lambda}\right)^2\ln
\left(\frac{\lambda}{\xi_p}\right)\approx 10^{7}
\left(\frac{m^*_\p/m_\p}{0.5}\right)^{-1}
\left(\frac{x_\p}{0.05}\right)\left(\frac{\rho_\n}{2\times
    10^{14} \mbox{g\, cm}^{-3}}\right) \mbox{ erg cm}^{-1},
\ee
where $m^*_\p$ is the effective mass of the protons,
$\xi_\p\approx 20$~fm the coherence length of a proton vortex and
$\lambda\approx 100$~fm is the London penetration depth of the
magnetic field.  Neutron vortices will thus be immersed in a tangle of
far more numerous flux tubes.

The interaction between neutron vortices and flux tubes changes the
energy of the system in two ways: there will be a change in
condensation energies, as the superfluid and superconducting cores
overlap, and also a contribution from the interaction between the
magnetic fields of the two vortices, as described in
Sec.~\ref{sec:mutial_friction}.  If the overlap reduces the energy of
the overall configuration neutron vortices are effectively `pinned' to
flux tubes, to some extent in the same way as they can be pinned to
ions in the crust.

The contribution due to the change in condensation energy is
\cite{1985PAZh...11..196M,1989ASIC..262..457S}
\be
\Delta E_c \approx 0.13\, \mbox{MeV} 
\left(\frac{\Delta_\p}{1\mbox{MeV}}\right) 
\left(\frac{x_\p}{0.05}\right)^{-1}
\left(\frac{m^*_\n/m_\n}{1}\right)^{-2}\left(\frac{m^*_\p/m_\p}{0.5}\right)^{-1}
\label{energia2}
\ee
with $m_\n^*$ the neutron effective mass, $\Delta_\p$ the proton
pairing gap, while the change in energy due to the magnetic
interaction \cite{1991MNRAS.253..279J, 1992ApJ...399..213C}:
\be
\Delta E_{\mg} = 2\frac{ B^i_\n B_i^\Phi}{8\pi} (\pi\lambda^2
l_\lambda),
\label{energiaB}
\ee
with $B^i_\n$ the magnetic field along the neutron vortex,
$B^i_\Phi\approx 10^{15}$ G that along the proton vortex, and
$l_\lambda$ is the overlap length between vortex and
flux-tube. Keeping in mind that neutron vortices can be considered
rigid on lengthscales approximately 100 times larger than those over
which a fluxtube can bend, we can average the expression in
(\ref{energiaB}) to obtain \cite{2012MNRAS.421.2682L}:
\be
\Delta E_{\mg}\approx 10
\left(\frac{m^*_\p/m_\p}{0.5}\right)^{-1/2}\left(\frac{|m_\p-m^*_\p|/m^*_\p}{0.5}\right)\left(\frac{x_\p}{0.05}\right)^{1/2}\left(\frac{\rho_\n}{2\times
    10^{14} 
\mbox{ g/cm$^3$}}\right)^{1/2} \mbox{ MeV}.
\label{energia1}
\ee
In general one may expect also a dependence on the inclination angle
of the global magnetic field with the rotation axis, and on vortex
tension, see Refs. \cite{2016MNRAS.462.1453G,2011Ap.....54..100S} for
a discussion of toroidal flux tubes and the effect of bending and
pinning in this case.  Nevertheless the above averaged expression
illustrates that the pinning force will be sizeable.  Pinned vortices
can thus `push' magnetic flux tubes, possibly winding up a strong
toroidal component of the magnetic field in magnetars
\cite{2011ApJ...740L..35G} and leading to the long term expulsion of
magnetic flux from the star as the vortex array expands while the star
spins down \cite{1990CSci...59...31S, 1998ApJ...492..267R,
  2000ApJ...532..514J, 2006MNRAS.365..339J}.

If, on the other hand, the pinning force cannot balance the Magnus
force, vortices are forced to cut through flux tubes. This process
will excite Kelvin waves along the vortex, leading to strong
dissipation and mutual friction \cite{1992ApJ...387..276E,
  2003PhRvL..91j1101L}.  The energy released at every vortex/flux tube
intersection is \cite{2003PhRvL..91j1101L}
\be \Delta
E_{vf}=\frac{2}{\pi} \frac{\Delta E_{\mg}^2}{\rho_\n \kappa \lambda}
(v_\lambda w_{\n\p})^{-1/2} ,
\ee
where $v_\lambda=\bar{h}/2m_K\lambda\approx 10^9$ cm/s is the
characteristic velocity of a Kelvon of effective mass $m_K$. The
energy loss rate per unit volume is thus
\be
\dot{\mathcal{E}}=\frac{n_\vv w_{\p\n}}{{l_{\phi}}^2} 
\Delta E_{vf}.
\label{loss} 
\ee 
By equating the expression in (\ref{loss}) to
the work done by the mutual friction force we can derive the drag
coefficient for vortex/flux tube cutting \cite{2014MNRAS.441.1662H}
\bea \mathcal{R} &=& \mathcal{R}_0
\left(\frac{v_\lambda}{w_{\p\n}}\right)^{3/2},\nonumber\\
\mathcal{R}_0 &=& \frac{2}{\pi}\left(\frac{\Delta E_{\mg}}{\rho_\n \kappa
    \lambda v_\lambda}\right)^2\approx 1.3\times 10^{-10}
\left(\frac{B}{10^{12}\mbox{G}}\right) .
\eea 
Note that the mutual friction coefficient is now velocity dependent,
and the lower the relative velocity the larger the friction. In
practice as soon as vortices start moving they are unlikely to be able
to continue and the system will move back towards the pinned state
\cite{2014MNRAS.441.1662H}.

\subsection{Pinning-repinning of vortices}
\label{sec:repinning}

Let us consider the forces acting on a massless vortex segment. If the
vortex is free the force balance equation, averaged over 
a number $n_\vv$ of vortices per unit area, is
\be
\kappa n_\vv \epsilon_{ijk} \hat{\Omega}^j (v^k_\n-v_\vv^k)
+\kappa n_\vv \mathcal{R} (v_i^\p-v_i^\vv)=0.
\label{free} 
\ee
One can solve Eq.~(\ref{free}) for the vortex velocity $v^i_\vv$ (the
direction of which which will depend on $\mathcal{R}$) and obtain the
standard form of mutual friction in Eq.~(\ref{MF}).  Pinning to the
ions in the crust or flux tubes in the core modifies the force balance
equation, because now some of the vortices may be immobilized by the
pinning.  When averaging over large number of vortices we can assume
that only a fraction $\gamma$ of them is free, leading to a force
balance equation of the form:
\be
\gamma \kappa n_\vv \epsilon_{ijk} \hat{\Omega}^j (v^k_\n-v_\vv^k)+\gamma \kappa n_\vv \mathcal{R} (v_i^\p-v_i^\vv)+(1-\gamma)\kappa n_\vv \epsilon_{ijk} \hat{\Omega}^j (v^k_\n-v_\p^k) + f_i^{\pin} =0
\label{pinned} 
\ee
where $f_i^{\pin}$ now balances the Magnus force on the
$(1-\gamma) n_\vv$ pinned vortices, for which we have assumed that
$v^i_{\vv} \vert_{\mbox{pinned}}=v_\p^i$. The force acting on the fluids is
thus $f_i^{\n, \pin}=-f_i^{\p, \pin}=f_i^{ \pin}$.

The quantity that is needed for the equations of motion in 
Eqs.~(\ref{euler}) and (\ref{pinned}) is thus the pinning force per unit
length acting on a vortex, which is highly uncertain. The pinning
force per pinning site can, in fact, be quite readily obtained
theoretically, as it depends only on the difference in energy between
the configuration where the vortex overlaps with an individual pinning
site, and that in which it is outside. Nevertheless even in this case
significant uncertainties remain, with different results in the
literature disagreeing also on whether the interaction is attractive
or repulsive, i.e. on whether one has pinning to nuclei or
interstitial pinning \cite{2002ApJ...569..381P,2004NuPhA.742..363D,
  2006PhLB..640...74D}. Note, however, that to understand the dynamics
of the fluid we are mainly interested in the magnitude of the pinning
force, and not in its sign.

For the case of pinning of neutron vortices to nuclei in the
crust the maximum pinning force acting on a vortex can be estimated
as \cite{1977ApJ...213..527A}
\be
|F_{\pin}|\approx
\left(n_{\rm out} \, E^{\rm cond.}_{\rm out} 
-n_{\rm in} \, E^{\rm cond.}_{\rm in} \right)
\frac{V}{\xi_n},
\label{oldpin}
\ee
where $V$ is the volume of a nuclear cluster and $\xi_n$ is the
coherence length of the nutron vortex, which defines the scale of the
interaction, `in' and `out' refer to quantities taken within the
nuclear cluster and in the free neutron gas,
$E^{\rm cond.}\simeq 3\Delta_n^2/8\ep_{Fn}$ is the condensation energy of
neutron fluid per unit volume.  The pinning force per
unit length depends however on the difference in energy between
different configurations of a vortex that encounters several pinning
sites and may bend to reduce its energy. It is thus, generally a
function of of the orientation of the vortex with respect to the
lattice
\be
f_{\pin}(\theta, \phi)=|F_{\pin}|\frac{\Delta n (\theta,\phi)}{l_T},
\label{unitl}
\ee
where the angles $(\theta, \phi)$ are taken with respect to a
reference axis and $l_T\approx 10^2-10^3 R_{WS}$ is the length scale
over which a vortex can bend, determined by the tension in
Eq.~(\ref{Tension}), and $R_{WS}$ is the radius of the Wigner Seitz
cell.  It was pointed out early on in Ref.~\cite{1991ApJ...373..208J}
that vortex rigidity plays an important role, as for an infinite
vortex all configurations would be energetically equivalent (i.e. they
would intercept the same number of pinning sites) and there would thus
be no pinning at all.

\begin{figure}[t]
\centerline{\includegraphics[width=10cm]{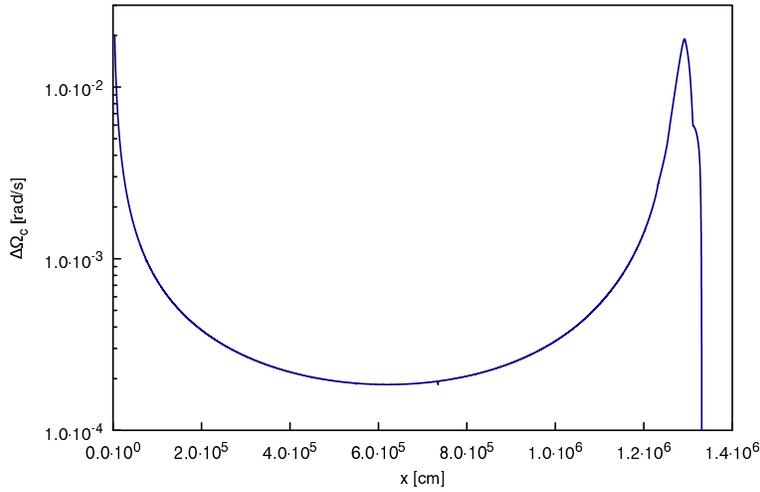}}
\caption{Critical lag $\Delta\Omega_c$ for unpinning for a
  $1.4 M_\odot$ neutron star described by the GM1 equation of state as
  described in \cite{2012MNRAS.427.1089S,2016MNRAS.455.3952S}.
}\label{lag}
\end{figure}

Recently calculations have been carried out for a realistic setup by
Ref.~\cite{2016MNRAS.455.3952S}, who averaged the expression in
Eq.~(\ref{unitl}) over all orientations of a vortex with respect to a
BCC lattice, and by Ref.~\cite{2016PhRvL.117w2701W} who studied the
interactions of a vortex with a pinning site in the time dependent
local density approximation.  These studies find that vortex tension
and bending is indeed fundamental for the physics of pinning. The
calculations differ in several aspects, the most notable of which is
that Ref.~\cite{2016PhRvL.117w2701W} obtains interstitial pinning
while Ref.~\cite{2016MNRAS.455.3952S} finds nuclear pinning, but the
authors do not include long range repulsive terms due to the Bernulli
force in the superfluid~\cite{2016MNRAS.462.1453G}. The angular
momentum reservoir is, however, independent of the sign of the pinning
force, and both sets of authors obtain pinning forces that are
dynamically significant and can explain the observed glitching
activity of the Vela pulsar. By simply balancing the Magnus force with
the pinning force Ref.~\cite{2016MNRAS.455.3952S} finds that velocity
differences up to $|w^{\p\n}|\approx 10^4$ cm s$^{-1}$ can be
sustained in the crust, which can explain the observed glitching
activity of the Vela and other pulsars, also in the presence of strong
entrainment \cite{2016MNRAS.455.3952S}. An example of the critical lag
profile in a neutron star, obtained for the pinning of
Ref.~\cite{2016MNRAS.455.3952S}, is shown in Fig.~\ref{lag}.

The results of dynamical simulations can also be used to study the
problem of repinning  of free vortices. This is a fundamental issue, as
while estimating when a pinned vortex will unpin allows us to estimate
how much angular momentum can be stored and released in a glitch,
calculating when a vortex will re-pin allows us to understand whether
vortices will `creep' out, gradually spinning down the star, or expel
vorticity in `avalanches'.

Macroscopic vortex dynamics in a spinning down container was studied
on the basis of Gross-Pitaevskii equations in
Refs.~\cite{2011MNRAS.415.1611W,2012PhRvB..85j4503W}, who have shown
that the main unpinning trigger for vortices is the proximity effect,
i.e. the change in Magnus force due to the motion of neighbouring
vortices, which can lead to forward or backward propagating vortex
avalanches. These can, in turn, trigger a glitch. Computational
limitations, however, constrain these simulations to a small number of
vortices (typically of the order of hundreds) separated by at the most
tens of pinning sites. This is in contrast with the situation
encountered in neutron stars where a large number
$N_\vv\gtrsim 10^{12}$ of vortices must move together in a
glitch. These are on average separated by a large number (of the order
of $10^{10}$) pinning sites. Nevertheless, the external spin-down
drives the system by increasing the lag between the superfluid and
normal fluid to the critical value for unpinning. It is thus crucial
to understand whether the system can self-adjust and hover close
enough to the critical lag that vortices can unpin and skip over many
pinning sites in order to knock on neighbouring vortices and allow an
avalanche to propagate.

\begin{figure}[t]
\centerline{\includegraphics[width=14cm]{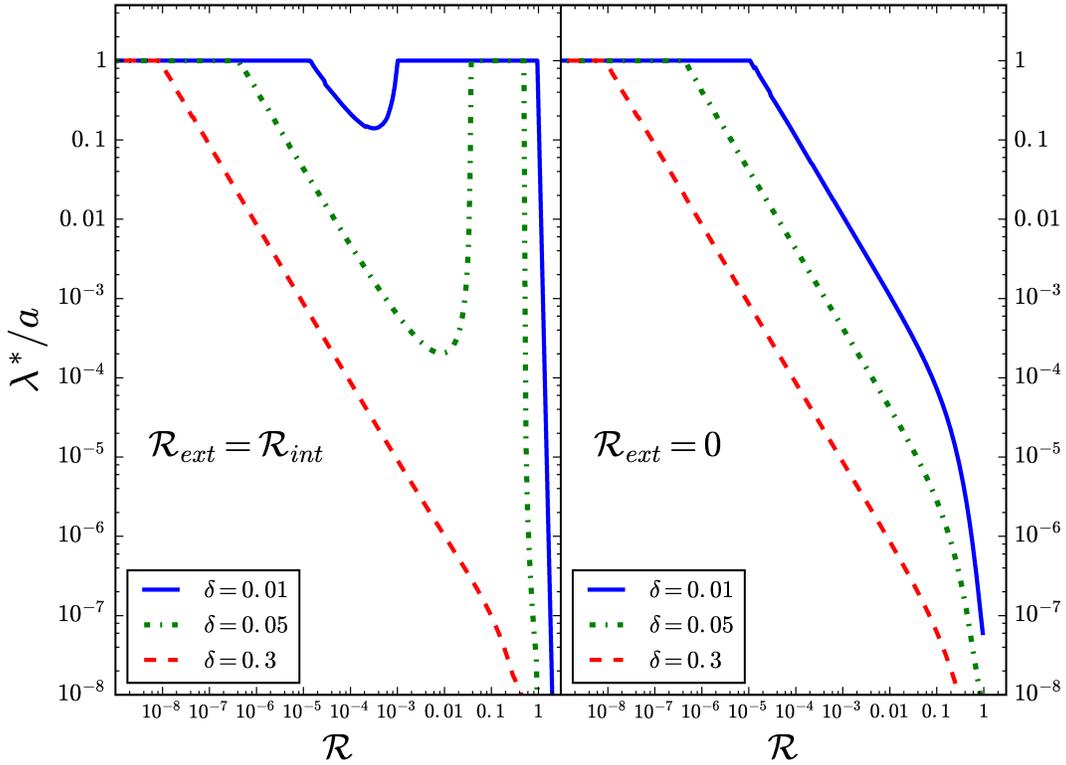}}
\caption{Mean free path $\lambda^*$ of a vortex, normalized to the
  intervortex spacing $a$, for different values of
  $\delta=(\Delta\Omega_c-\Delta\Omega)/\Delta\Omega_c$, with
  $\Delta\Omega_c$ the critical lag for unpinning, as described in
  \cite{2016MNRAS.461.2200H}. The left pannel shows the case in which
  mutual friction is described by the same parameter $\mathcal{R}$
  inside and outside the pinning potential, while the right panel is
  for the case in which there is no mutual friction outside the
  pinning potential. In general avalanches can propagate if mutual
  friction is weak, and especially in the case in which mutual
  friction is the same everywhere, if $\delta\lesssim 0.05$, i.e. the
  system is close to the critical lag for unpinning.}\label{repin}
\end{figure}

The problem of vortex re-pinning was investigated in Refs.
\cite{1995MNRAS.277..225S,2016MNRAS.461.2200H}, by considering vortex
motion in a parabolic pinning potential. Pinning of a moving vortex in
a random potential was also studied in
Ref.~\cite{2009PhRvL.102m1101L}. In particular
Ref.~\cite{2016MNRAS.461.2200H} calculated the mean-free path of a
straight vortex for scattering off cylindrical pinning sites, and
found that the main parameters that control repinning are the strength
of the mutual friction and, crucially, how close the system is to the
critical threshold lag for unpinning. From Fig.~\ref{repin} we can
deduce that if the system is within 5\% of the critical lag for
unpinning, a vortex can move a distance comparable to the inter-vortex
separation and knock on other vortices, causing an avalanche, for
realistic values of the mutual friction. Studies by
Refs.~\cite{2001PhRvL..86.1785C,2003PhRvL..91j7002M,2010PhRvB..82m4519F}
have also shown that the geometry of the lattice can play a crucial
role, with a more disordered lattice behaving like a plastic system,
in which unpinned and pinned vortices coexist, and an ordered lattice
behaving like an elastic system in which there is a sharp transition
to mass unpinning.

Before moving on, let us note that if protons form a type II
superconductor in the core, as was discussed in
Sec.~\ref{sec:fluxtubes}, the energy cost of vortex/flux-tube cutting
can lead to pinning. A simple estimate of the pinning force per unit
length $f^{\pin,\phi}$ can be obtained from the expression for the
overlap energy in (\ref{energiaB})
\be
f^{\pin,\phi}\approx \frac{E_{\mg}}{\lambda l_\phi},
\label{pincore}
\ee
where $\lambda$ is the London penetration length, $l_\phi$ the
distance between fluxtubes and $E_{\mg}$ is defined in
Eq.~(\ref{energia1}). The force (\ref{pincore}) is balanced by the
Magnus force for a critical velocity
$|w^{\p\n}|\approx 5\times 10^3 (B/10^{12}\mbox{G})^{1/2}$ cm
s$^{-1}$, which indicates that this force could play an important role
in a glitching model based on unpinning of vortices in the core.
Note, however, that the estimate \eqref{pincore} does not account for
the effect of averaging over different orientations of the vortex
lattice, and is thus an upper limit on the pinning force.

\section{Macrophysics of superfluidity in neutron stars}
\label{sec:macro_super}

\subsection{Glitches and post-glitch relaxations}
\label{sec:glitch}

Pulsar glitches are sudden spin up events in the otherwise steadily
decreasing rotational frequency of pulsars. These were observed in the
Vela pulsar soon after the discovery of radio pulsars
\cite{1969Natur.222..229R,1969Natur.222..228R}.  The initial jump in
frequency in the case of the Vela pulsar is instantaneous to the
accuracy of the data (with the best upper limit of $\tau_r\lesssim 40$
s coming from the Vela 2000 glitch \cite{2002ApJ...564L..85D}), but is
often accompanied by an increase in spin-down rate that relaxes back
towards the pre-glitch values on longer timescales (ranging from
minutes to months).  The long-time scales of relaxations following
glitches were taken as an evidence for the presence of a loosely
coupled superfluid component in the star \cite{1969Natur.224..872B}.

Initial studies attributed the long relaxation times to the slow
coupling of the core of the star due to the weak coupling of vortices
to the electron fluid according to Eq.~\eqref{eq:np_relax},
see~Refs.~\cite{1969Natur.224..872B,1971PhRvD...4.1589F}.  Anderson
and Itoh \cite{1975Natur.256...25A} put forward the hypothesis that
the glitches are linked to a pinned superfluid in the star, that is
decoupled from the observable `normal' component, and whose sudden
re-coupling (due to unpinning) leads to an exchange of angular
momentum and a glitch \cite{1975Natur.256...25A}.

Following the idea of a pinned superfluid in the crust of a neutron
stars~\cite{1975Natur.256...25A} the initial models of glitches and
post-glitch relaxation concentrated on the detailes of the physics of
vortex pinning and unpinning mainly  in the crust and the fits of the models
to the observed behaviour of the Vela
pulsar~\cite{1984ApJ...276..325A,1984ApJ...278..791A,1985ApJ...288..191A,1988ApJ...330..835C}. The
sudden unpinning of neutron vortices was attributed to the a glitch
and their slow relaxation via thermal creep against pinning barriers
as the model of post-glitch relaxation.  The `creeping' velocity of
neutron vortices is given in these models by
\be v_r\approx v_0 \exp(-E_a/k_BT),
\label{vortici} 
\ee 
where $v_0\approx 10^7$ cm s$^{-1}$
\cite{1984ApJ...276..325A}, $k_B$ is Boltzmann's constant and $E_a$ is
the activation energy for unpinning \cite{1991ApJ...373..592L}. This
latter quantity in a first approximation can be taken as 
$E_a\approx E_p\left(1-{\Delta\Omega}/{\Delta\Omega_c}\right),$
with $E_p$ the pinning energy, $\Delta\Omega$ the lag between the
superfluid neutrons and the crust, and $\Delta\Omega_c$ the critical
lag for unpinning.  The equations of motion for the frequency of the
observable `normal' component of the star $\Omega_\p$ are
\be
I_\p\dot{\Omega}_\p=N_{\rm ext}+\sum_i I_\n^i \frac{2\Omega_\n^i}{\varpi}
v_r^i ,
\label{Creep} 
\ee 
where $I_\p$ is the moment of inertia of the crust and all components
tightly coupled to it, $\varpi$ the cylindrical radius, $N_{\rm ext}$
is the external torque, and the superfluid is divided into a number
$i$ of different regions, with associated moment of inertia $I_n^i$,
angular velocity $\Omega_\n^i$ and vortex velocity $v_r^i$, calculated
from (\ref{vortici}).  The solutions to Eqs.~\eqref{Creep} admit two
regimes. If the steady state lag $\Delta\Omega$ is much smaller that
$\Delta\Omega_c$ then the response of the system is linear and the
region contributes to the frequency relaxation exponentially after a
glitch \cite{1993ApJ...409..345A}. This is also the regime that can be
modelled in terms of mutual friction coupling due to a small number of
free vortices
\cite{2010MNRAS.409.1253V,2012MNRAS.420..658H,2014ApJ...789..142V,2016MNRAS.460.1201H}. If
$\Delta\Omega\approx\Delta\Omega_c$ the response will be nonlinear,
and the contribution of the region to the relaxation takes the form of
a Fermi function
\cite{1984ApJ...276..325A,2014ApJ...789..141L}. Recent work has also
shown that the non-linear response of creep to glitches can be used to
interpret the inter-glitch behaviour of the Vela pulsar, and predict
the occurrence of the next glitch~\cite{2016arXiv161203805A}.  The
creep model has been more recently extended to the scenario where
neutron vortices creep against the core
flux-tubes~\cite{2014ApJ...789..141L,2016MNRAS.462.1453G}, with
essentially the same physical picture of post-jump relaxation involved.
Pinning of superfluid neutron vortices to proton vortices in the core
naturally extends the reservoir of angular momentum available for a
glitch thus potentially explaining the observed activity of the Vela
pulsar~\cite{2014ApJ...788L..11G}, although a large amount of pinned
vorticity in the core is not consistent with linear models for the
recovery of Vela glitches \cite{2013ApJ...764L..25H}.  The original
crustal vortex creep model relied on the assumption, derived from the
short relaxation times found in Ref.~\cite{1984ApJ...282..533A}, that
the core is coupled to the crust on short dynamical time-scales, which
are unobservable in glitches and their relaxations.

An alternative to crust-based models is the vortex cluster model of
dynamics of superfluid neutron vortices in the core of a neutron star,
where the coupling between the superfluid and normal component occurs
on much longer time-scales~\cite{1995ApJ...447..305S}. Models of
pulsar glitches and post-glitch relaxations based on the superfluid
core rotation in the absence of pinning between the neutron vortices
and flux-tubes were developed within the vortex cluster model and
applied to Vela glitches in Ref.~\cite{1995ApJ...447..324S}. The
glitch itself can arise in these models through the interaction of
vortex clusters with the crust-core
interface~\cite{1999MNRAS.307..365S}; the core moment of inertia alone
was estimated to be sufficient to account for both glitches and
post-glitch relaxations~\cite{1995ApJ...447..324S,1999MNRAS.307..365S}.

\begin{figure}[t]
\centerline{\includegraphics[width=10cm]{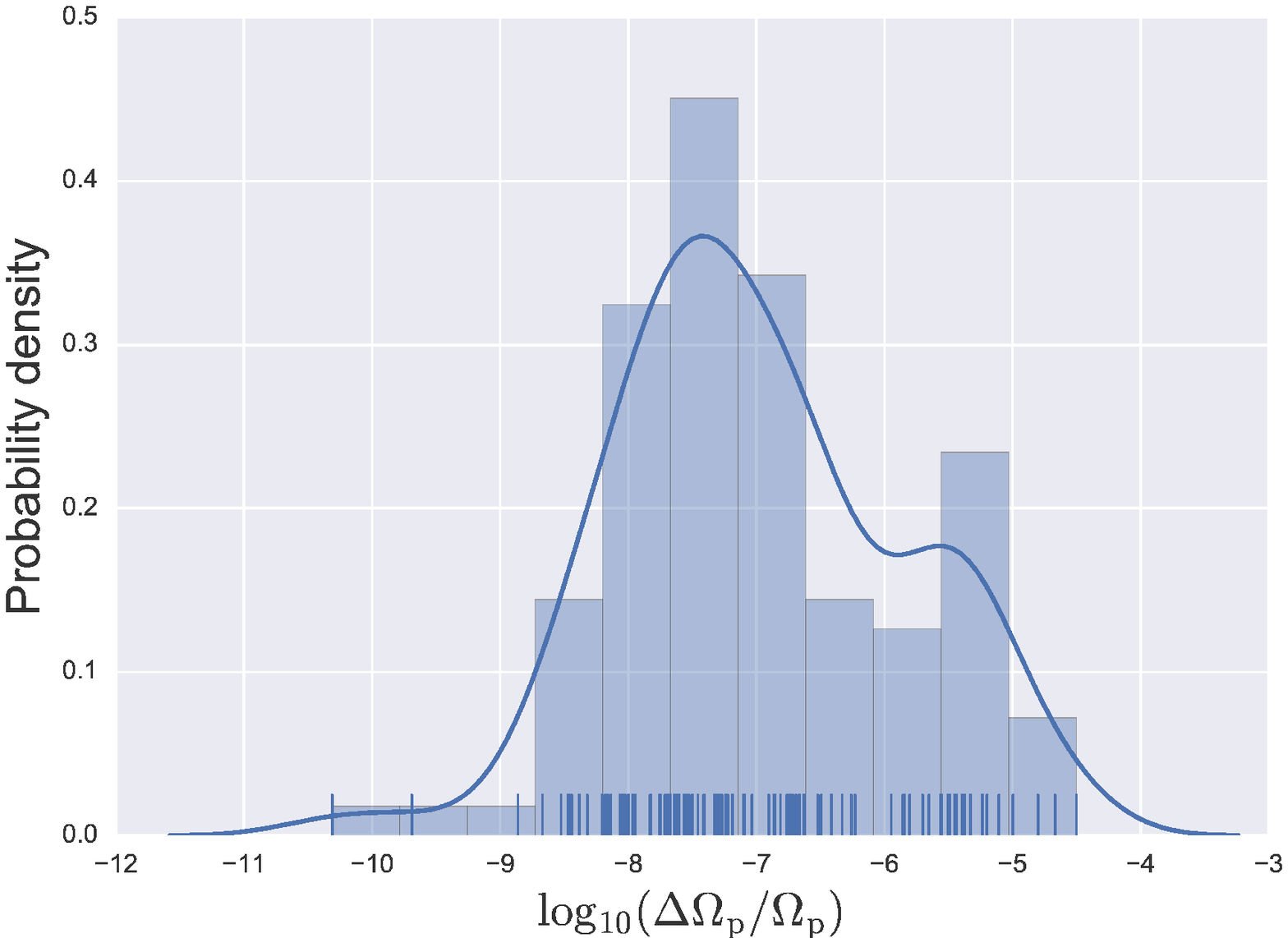}
\includegraphics[width=10cm]{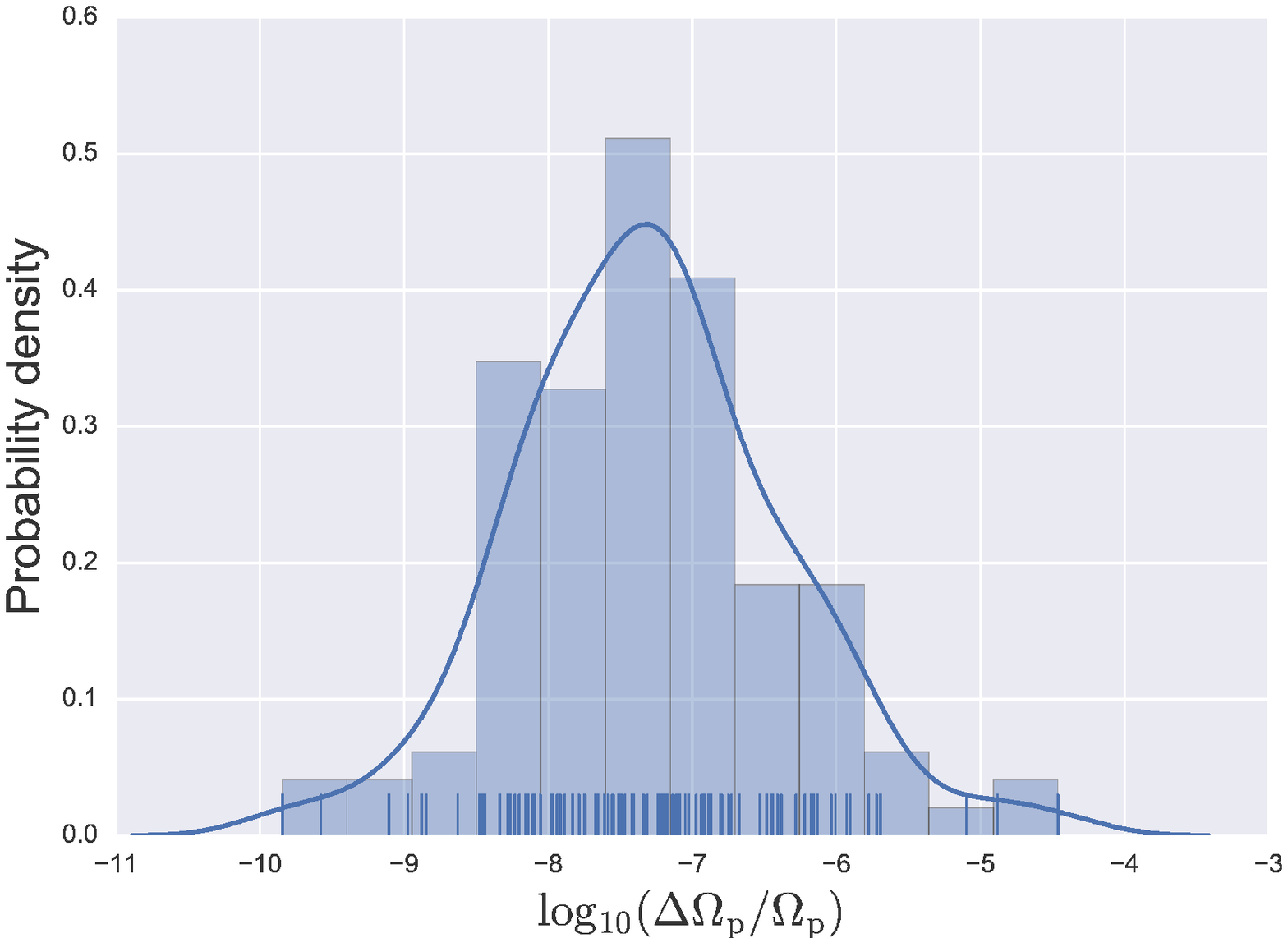}}
\caption{Probability distribution function for glitch sizes
  $\Delta\Omega_\p/\Omega_\p$ for a microscopic waiting time $t_w=0.1$
  days and two different microscopic power law indices: $n=-1.05$
  (left) and $n=-1.5$, as described in Ref.~\cite{2016MNRAS.461L..77H}. In
  both cases there is a clear deviation from a power-law, with the
  appearance of a cutoff at low sizes, as observed in the Crab pulsar
  \cite{2014MNRAS.440.2755E}. }\label{distro1}
\end{figure}

If the system hovers close to the critical unpinning threshold vortex,
avalanches may propagate in the neutron star
interior~\cite{2016MNRAS.461.2200H}, which are thus a viable mechanism
for triggering a glitch
\cite{1988ApJ...330..835C,2012PhRvB..85j4503W}. A glitching pulsar
would thus behave as a self organised critical system, in which slowly
increasing global stresses (due to the external spin-down torque that
drives the increase in lag and thus Magnus force) are released rapidly
and locally via nearest neighbour interactions between vortices. Such a
system is scale invariant and one expects the distribution of sizes of
the avalanches to be a power-law, and the distribution of waiting
times an exponential. This is generally what is observed in the pulsar
population \cite{2008ApJ...672.1103M}, with the notable exception of
the Vela pulsar and PSR J0537-6910, which exhibit a quasi-periodicity
in their glitching behaviour and for which the glitches can be
predicted \cite{2006ApJ...652.1531M,2016arXiv161203805A}. This
behaviour in the case of PSR J0537 is generally attributed to the
presence of crust-quakes, but may also be the consequence of the
timescale of the external driving (the spin-down) being short compared
to the time-scale on which stress is released locally by the vortices.
Another system in which the distribution of glitch sizes appears to
deviate from a power-law is the Crab pulsar \cite{2014MNRAS.440.2755E}
for which there appears to be a cut off for small glitch sizes. This
behaviour is, however, natural if one considers not only the exchange
of angular momentum due to vortex motion, but also the coupling
timescale due to mutual friction. Ref.~\cite{2016MNRAS.461L..77H}
investigated this by considering a multifluid system described by
Eq.~\eqref{euler}, in which however only a number $\gamma n_\vv$ of
vortices is free at any given time, with $\gamma\leq 1$. The value of
$\gamma$ is randomly drawn from a power-law distribution after a
waiting time $t_w$ (randomly drawn from an exponential distribution)
and the results of these simulations show that small values of
$\gamma$ not only correspond to a small amount of angular momentum,
but also to an effective reduction in the average mutual friction and
an increased coupling timescale. In this case the event does not
appear as a sudden jump in frequency, but is more gradual and closer
to timing noise, thus not being recognised as a glitch by detection
algorithms and producing a cutoff in the size distribution for small
glitches, see Fig.~\ref{distro1}.

In addition to the mechanisms mentioned above there exist a number of
other candidate mechanism for glitches. As already mentioned crust
quakes may drive glitches
\cite{1969Natur.223..597R,1976ApJ...203..213R, 2000A&A...361..795C} as
has been suggested for PSR J0537-6910 and also the Crab pulsar
\cite{2006ApJ...652.1531M}, and hydrodynamical instabilities may also
lead to a glitch \cite{2005MNRAS.361..927M,2009PhRvL.102n1101G}.

In the above discussion of the pinning-based trigger mechanisms for a
glitches we tacitly assumed that pinning occurs between vortices
and ions in the crust or vortices and flux-tubes in the core. The
extent and location of the pinning region can, however, be studied
more quantitatively by examining both the size of the maximum glitch
recorded in a pulsar \cite{2011ApJ...743L..20P, 2012MNRAS.427.1089S},
and its `activity' $\mathcal{A}$, i.e the amount of spin-down that is
reversed by glitches during observations
\be
\mathcal{A}=\frac{1}{t_{\rm obs}}\sum_i\frac{\Delta\Omega_\p^i}{\Omega_\p},
\ee
where the sum is performed over all recorded glitches in a time
$t_{\rm obs}$ and $\Delta\Omega_\p^i$ is the recorded size of glitch
$i$. From angular momentum conservation over a glitch one has that the
ratio between the moment of inertia of the superfluid reservoir $I_\n$
and that of the `normal' component $I_\p$ is
\be
\frac{I_\n}{I_\p}\approx -\frac{\Omega_\p}{\dot{\Omega}_\p} \mathcal{A} (1-\varepsilon_\n)
\ee
Refs.~\cite{2012PhRvL.109x1103A,2013PhRvL.110a1101C} noted that in the
presence of strong neutron entrainment $\varepsilon_\n$, such as is
predicted in the crust where $\varepsilon_\n\approx 10$ due to Bragg
scattering \cite{2012PhRvC..85c5801C}, the crust cannot store enough
angular momentum to explain the observed activity of the Vela pulsar
(unless the star has a very small mass $M\lesssim 1 M_\odot$). The
core must be involved in the glitch mechanism.  The neutron superfluid
is, in fact, expected to extend into the core and models that extend
the reservoir beyond the crust can predict the observed activity of
the Vela and other pulsars
\cite{2015MNRAS.454.4400N,2015SciA....1E0578H}. The observed activity
of a pulsar, together with the size of its maximum glitch, can then
potentially be used to determine the mass of a glitching pulsar and
constrain the equation of state of dense
matter~\cite{2015SciA....1E0578H,2017NatAs...1E.134P}.

Finally let us note that the effect of both classical and superfluid
turbulence have been ignored in the above discussion, but my have an
impact on glitch physics. Transitions between turbulent and laminar
regimes could explain the short spin up timescales and long
inter-glitch timescales \cite{2006ApJ...651.1079P,2007ApJ...662L..99M}
and shear-driven turbulence can contribute to low frequency
fluctuations in the spin of the star, i.e. so-called `timing noise'
\cite{2014MNRAS.437...21M}.

\subsection{Oscillations in superfluid stars}

Neutron stars are expected to be prolific emitters of gravitational
waves \cite{2015ASSP...40...85H, 2015PASA...32...34L} and in particular there are several
modes of oscillation of the star that could lead to detectable
emission. In particular the most promising modes for ground based
detection are the $f$-mode, or fundamental mode, and the $r$-mode,
analogous to Rossby waves in the ocean, which can be driven unstable
due to gravitational wave emission and grow to large amplitudes
\cite{1998ApJ...502..708A,1998ApJ...502..714F}.  In order to assess
the detectability of these signals it is thus crucial to understand in
which region of parameter space the modes can be driven unstable by
gravitational wave emission, and in which region, on the other hand,
they are rapidly damped by viscosity.  At high temperatures
($T\gtrsim10^9$ K) bulk viscosity is the main damping mechanisms and
matter is not expected to be superfluid. At lower temperatures,
however, superfluidity has a strong impact on the damping. On the one
side superfluidity leads to a suppression in the neutron-neutron and
neutron-proton scattering events that gives rise to shear viscosity,
which in this case is mainly due to electron-electron scattering and
reduced with respect to the case in which neutrons are normal
\cite{2005NuPhA.763..212A}. On the other superfluidity opens a new
dissipative channel by allowing for vortex mediated mutual friction. 

It has been established early on that the doubling of the degrees of
freedom in the superfluid component doubles the number of oscillations
modes; these have been studied in Newtonian
theory~\cite{1994ApJ...421..689L,1995A&A...303..515L,2001PhRvD..63b4016S}
and general relativistic setting~\cite{1999PhRvD..60j4025C}.  The
$l\le 2$ modes of incompressible superfluid self-gravitating fluids
(both axially symmetric and tri-axial) were studied in
Refs.~\cite{2001PhRvD..63b4016S,2001LNP...578...97S} in the presence
of mutual friction and viscosity using the tensor virial method. In
the absence of  shear viscosity the mutual friction can be
eliminated from the equations describing the center-of-mass motions of
the two fluids and it acts to damp only the relative motions of the
two fluids. Shear viscosity which acts only in the normal fluid
breaks the symmetry of the Euler equations for the normal fluid and
superfluid and, therefore, couples these two sets of
modes~\cite{2001PhRvD..63b4016S,2001LNP...578...97S}.  These initial
studies were followed by studies which obtained the analogues of the
$f$ and $p$ modes in superfluid neutron
stars~\cite{2002A&A...393..949P} and included the general relativity
in the mode description in the case of non-rotating stars
\cite{2002PhRvD..66j4002A,2014PhRvD..90b4010G}.  Furthermore, much
work has been concentrated on the $r$-modes of the superfluid neutron
stars~\cite{2003ApJ...586..403L,2003PhRvD..67l4019Y,2003MNRAS.344..207Y,2004MNRAS.348..625P,2009MNRAS.397.1464H,2010PhRvD..82b3007A}
which may play a key role in the dynamics and stability of rapidly
rotating neutron stars. Progress has been made in understanding the
oscillations modes at finite
temperature~\cite{2006MNRAS.372.1776G,2011PhRvD..83j3008K,2013CQGra..30w5025A}
as well as the influence of crust elasticity~\cite{2012MNRAS.419..638P}.

Let us examine this problem in detail. The linearised version of the
equations of motion in (\ref{euler}) can be written in terms of two
sets of degrees of freedom, one that represents the `total' motion of
the fluid (and would be present also in a normal, non-superfluid,
star), and one that represents the counter-moving motion.  Let
us consider the case in which mutual friction is the only dissipative
mechanism acting on the system. By introducing the total mass flux
\be \rho\delta
v^k=\rho_\n\delta v_\n^j + \rho_\p \delta v_\p^j 
\ee 
with
$\rho=\rho_\n+\rho_\p$ and where $\delta$ represents Eulerian
perturbations, we can write a `total' Euler equation for the total
velocity $v^i$, in a frame
rotating with angular velocity $\Omega$, 
\be
\partial_t\delta v_i+\nabla_i\delta\Phi+\frac{1}{\rho}\nabla_i\delta p -\frac{1}{\rho^2}\delta\rho\nabla_i p +2\epsilon_{ijk}\Omega^j\delta v^k=0
\ee
where the pressure $p$ is obtained from
\be
\nabla_i p=\rho_\n \nabla_i\tilde{\mu}_\n+\rho_\p
\nabla_i\tilde{\mu}_\p .
\ee
We also have the standard continuity equation
\be
\partial_t\delta\rho+\nabla_j (\rho\delta v^j)=0 .
\ee
As already mentioned these are identical to the perturbed equations of
motion for a single fluid system, and quite notably the mutual
friction term drops out of the equations. The mutual friction
naturally appears in the equations of motion for the second degree of
freedom, which we can write in terms of the perturbed relative
velocity $\delta w^j=\delta v_\p^j-\delta v_\n^j$. The `difference'
Euler equation is
\be (1-\varepsilon_\n x_\p^{-1})\partial_t\delta w_i +
\nabla_i\delta\beta+2\tilde{\mathcal{B}}^{'}\epsilon_{ijk}\Omega^j\delta
w^k -
\tilde{\mathcal{B}}\epsilon_{ijk}\hat{\Omega}^j\epsilon^{klm}\Omega_l\delta
w_m=0, \ee
where we have defined the local deviation from chemical equilibrium
\be
\delta\beta=\delta\tilde{\mu}_\p-\delta\tilde{\mu}_\n
\ee
and $\tilde{\mathcal{B}}^{'}=1-\mathcal{B}^{'}/x_\p$ and
$\tilde{\mathcal{B}}=\mathcal{B}/x_\p$ with $x_\p=\rho_\p/\rho$ the
proton fraction. The continuity equation for the proton
 fraction is:
\be
\partial_t\delta x_\p +\frac{1}{\rho}[x_\p(1-x_\p)\rho\delta w^j]+\delta v^j\nabla_j x_\p=0
\ee
The degrees of freedom are thus explicitly coupled unless $x_\p$ is constant. In general we do not expect to find any mode of oscillation in a realistic neutron star that is purely co-moving, and thus all modes are affected, to some extent, by mutual friction.

We can define a conserved energy for the system by first defining a
`kinetic' term as an integral over a volume $V$
\be
E_k=\frac{1}{2}\int \left[ |\delta v|^2+(1-\varepsilon_\n/x_\p)x_\p(1-x_\p)|\delta w|^2\right]\rho\;dV
\ee
and a `potential' term
\be
E_p=\frac{1}{2}\int\left\{\rho\left(\frac{\partial\rho}{\partial
      p}\right)_\beta|\delta
  h|^2+\left(\frac{\partial\rho}{\partial\beta}\right)_p [2{\rm Re}(\delta h\delta\beta^*)+|\delta\beta|^2]-\frac{1}{4\pi G}|\nabla_i\delta\Phi|^2\right\}dV,
\ee
where a star represents complex conjugation. The perturbed equations of motion we have written down explicitly
include dissipative terms due to mutual friction. It is, however,
instructive to consider the non dissipative case and ignore the
contribution due to mutual friction. In this case
$\partial_t(E_k+E_p)=0$ and one can solve the problem for a mode with
time dependence $e^{i\omega t}$. If damping is weak, and procedes on a
timescale $\tau$ much longer than the period of the mode, i.e. one has
$\omega=\omega_r+i/\tau$, with $\tau>>1/\omega$, we can use the
solution of the non-dissipative problem to estimate the damping
timescale as
\be
\tau=\left|\frac{2(E_p+E_k)}{\partial_t E}\right|,
\label{tempo}
\ee
where $\partial_t E$ is obtained from a dissipation integral, in which
the dissipative terms due to viscosity are evaluated using the non
dissipative solution. In the case of mutual friction one can see that
this takes the form \cite{2009PhRvD..79j3009A}
\be
\partial_t E_\mathcal{B}=-2\int \rho_\n\mathcal{B}\Omega[\delta^m_i-\hat{\Omega}^m\hat{\Omega}^i]\delta w^{i*}\delta w_m\;dV.
\ee
If the damping timescale is sufficiently short, the estimate of $\tau$
in (\ref{tempo}) matches that obtained from the full mode
solution. Often, however, the full solution ot the dissipative problem
is not available, and (\ref{tempo}) is the only way to asses the
impact of viscosity.

Finally one must calculate the timescale for gravitational waves to
drive the mode. The energy lost to gravitational waves can be obtained
from a multipole expansion \cite{1980RvMP...52..299T}
\be
\partial_t E_{gw}=-\omega_r\sum_l N_l \omega_i^{2l+1}(|\delta D_{lm}|^2+|\delta J_{lm}|^2),
\ee
where $\omega_i$ is the frequency of the mode in the intertial frame
$N_l=(4\pi G)(l+1)(l+2)/\left\{c^{2l+1}l(l-1)[(2l+1)!!]^2\right\}$ and
the mass multipoles are
\be
\delta D_{lm}\approx \int\delta T_{00}Y^*_{lm} r^l dV
\ee
and the current multipoles are:
\be
\delta J_{lm}\approx-\int\delta T_{0j}Y^{B*}_{j,lm} dV,
\ee
with $Y^B_{j,lm}$ the magnetic multipoles \cite{1980RvMP...52..299T}
and $T_{\alpha\beta}$ is the two-fluid stress energy tensor defined in section (\ref{relativity}).
We refer the interested reader to \cite{2001IJMPD..10..381A,
  2009PhRvD..79j3009A} for a detailed discussion of the calculation,
and simply point out that in general
\beq
T_{00}&\approx& \delta \rho,\\
T_{0j}&\approx&\rho\delta v_j+\delta\rho v_j,
\eeq
i.e., only the co-moving degree of freedom radiates gravitationally to
leading order in rotation.

We are now potentially equipped to calculate the driving timescale
$\tau_{gw}$ and compare it to the mutual friction damping timescale
$\tau_\mathcal{B}$ for realistic neutron star modes. Clearly if
$\tau_\mathcal{B}\lesssim \tau_{gw}$ mutual friction will damp the
mode faster than gravitational radiation can drive it, and suppress
the instability, while in the opposite case a mode can grow to large
amplitudes and radiate gravitationally.

Let us first of all examine the fundamental, or $f$-mode. This is
essentially a `surface' mode for which the frequency
$\omega_f\sim \bar{\rho}^{1/2}$, with $\bar{\rho}$ the average density
of the star. In the case of the $f$-mode both Newtonian and
Relativistic studies have shown that mutual friction completely
suppresses the gravitational-wave driven instability below the
superfluid transition temperature \cite{2009PhRvD..79j3009A,
  2011PhRvL.107j1102G}. It may thus play a role in hot newly born
neutron stars \cite{2016PhRvD..94b4053P,2016A&A...586A..86S} but is
unlikely to be active in older, colder stars.

The situation is different for the $r$-mode. To first order in
rotation this mode is purely axial and comoving, leading to a single
multipole solution of the form:
\be
\delta v^j=\left[\sum_l \left(\frac{m}{r^2\sin\theta}U_lY_l^m \hat{e}^j_\theta+\frac{i}{r^2\sin\theta} U_l \partial_\theta Y_l^m \hat{e}^j_\phi\right)\right]\exp{(i\omega_0 t)}
\ee
where $U_l=Ar^{l+1}$ with $A$ a constant, $\omega_0={2m}/{l(l+1)}$ is
the frequency of the mode in the rotating frame, and we are using
spherical coordinates, with $\hat{e}^j_\theta$ and $\hat{e}^j_\phi$
unit vectors and $Y_l^m$ are spherical harmonics. The $l=m=2$ $r$-mode
thus provides the strongest contribution to gravitational wave
emission, as in a single fluid star. However, as we have mentioned,
the co-moving motion couples to the counter-moving degrees of freedom
at higher orders in rotation, leading to mutual friction damping
\cite{2009MNRAS.396..951P}.

For the $r$-mode standard mutual friction due to electron scattering
of vortex cores has little effect on the instability
\cite{2000PhRvD..61j4003L,2009MNRAS.397.1464H}. Strong mutual
friction due to vortex/flux-tube cutting in the core can, however,
have a strong impact on the instability and damp it in a large section
of parameter space. Furthermore strong dissipation due to vortex-flux
tube interactions limits the growth of the mode and sets an effective
saturation amplitude for it that may be smaller that the saturation
amplitude due to non-linear couplings to other modes
\cite{2014MNRAS.441.1662H}.

Furthermore \cite{2003ApJ...586..403L, 2014PhRvL.112o1101G,
  2014PhRvD..90f3001G, 2017arXiv170506027K} have suggested that for specific temperatures
there can be avoided crossings between the superfluid $r$-modes and
other inertial modes, leading to enhanced mutual friction
damping. This is in an interesting scenario as it would reconcile our
understanding of the $r$-mode instability in superfluid neutron star
with observational data on spins and temperatures of neutron stars in
Low Mass X-ray Binaries
\cite{2011PhRvL.107j1101H,2012MNRAS.424...93H,2013ApJ...773..140M},
and also predicts the presence of hot, rapidly rotating neutron stars
\cite{2014MNRAS.445..385C, 2016MNRAS.455..739K}.

Finally \cite{2009PhRvL.102n1101G,2013PhRvD..87f3007A} have pointed
out that in the presence of pinned vorticity the $r$-mode can grow
unstable if a large lag develops between the superfluid and the crust,
also possibly triggering a glitch.

\subsection{Long-term variabilities}

In addition of the phenomena discussed above neutron stars exhibit
long-term variability, which has been attributed to free
precession~\cite{2000Natur.406..484S,2016MNRAS.455.1845K}.
Theoretical studies of precession in neutron stars containing a
superfluid component indicate that the analogue of the free precession
in classical systems will be damped if the superfluid is coupled
strongly to the normal
component~\cite{1977ApJ...214..251S,1999ApJ...524..341S}. In addition
to this analogue of the classical precession, a fast precession mode
appears, which is associated with the doubling of the degrees of
freedom. To the leading order it is independent of the deformation of
the star and scales as the ratio of the moments of inertia of the
superfluid and normal components.  Nevertheless, free precession of
neutron stars has been modelled and applied to the available data with
some success~\cite{2001MNRAS.324..811J,2003MNRAS.341.1020W,
  2006MNRAS.365..653A,2012MNRAS.420.2325J,2017MNRAS.467..164A}, which
might be an evidence for the weak coupling of the superfluid to the
normal component. Long term variabilities in neutron stars can
also be understood in terms of the Tkachenko waves - oscillations of
the vortex lattice in the neutron
superfluid~\cite{1970Natur.225..619R}. The corresponding modes have
been studied in a number of setting including the damping by mutual
friction and shear viscosity in
Refs.~\cite{2008PhRvD..77b3008N,2009Ap.....52..151S,2011PhRvD..83d3006H,2011Ap.....54..111S}
and have been shown to be in the range relevant for the long-term
periodicities observed in pulsars.

\section{Conclusions and future directions}
\label{sec:conclusions}

This chapter provided an educational introduction to the physics of
superfluidity and superconductivity as well as a discussion of
selected subjects of current interest. Some basic aspects of the
physics of superfluidity in neutron stars are now well established:
the rough magnitudes of the pairing gaps, the existence of vortices
and flux tubes, the basic channels of mutual friction and mechanisms
of vortex pinning. The basic contours of the macro-physics behaviour
in glitches, their relaxations and other anomalies such as precession
and oscillations are also emerging. However, there are significant
uncertainties related to the details of the theory. For example, it is
a matter of debate whether the glitches occur in the crust, in the
core or in both components of the star. The same applies to the
post-glitch response of various superfluid shells that respond to a
glitch on observable dynamical time-scales. Finally, there are
phenomena that are not understood well at the basic level, such as for
example, the possibility of free precession in neutron stars.

One of the main difficulties in modelling these phenomena lies in the large
separation in scales that exists in neutron stars between the
interactions of vortices, flux tubes and clusters on the Fermi scale,
vortex-vortex interactions on the scale of millimeters and the large
scale hydrodynamics of the star. Future efforts
must thus focus on bridging the gap between these scales, both from a
theoretical and a computational point of view.

Future theoretical developments in the field will definitely
obtain impetus from observational programs in radio, X-ray and
gravitational wave astronomy. The SKA radio observatory, to become
operational in the upcoming decade, has the potential of significant
discoveries in pulsars astrophysics through the strong increase of the
number of observed pulsars (up to 30.000) and its high
sensitivity. The currently operating gravitational wave observatories
are sensitive probes of continuous gravitational wave radiation from
the pulsars and current upper limits on such radiation are sensitive
enough to constraint some physics of neutron star interiors (e.g., the
rigidity of the crust). In the future, some of the transients seen in
the electromagnetic spectrum may become observable through
gravitational waves, thus providing complementary information on their
dynamics. Finally, current and future X-ray observations of
pulsars are expected to put stronger constraints on the thermal
evolution models of neutron stars and their gross parameters (in
particular, neutron star radii will be measured by the NICER
experiment to a high precision).  Combined these `multi-messenger'
observations of neutron stars will provide us with a deeper
theoretical understanding of the workings of superfluids in neutron star
interiors.

\section*{Acknowledgements} 

B.H.  has received funding from the European Union’s Horizon 2020
research and innovation programme under grant agreement No. 702713.
A.S.\ is supported by the Deutsche Forschungsgemeinschaft (Grant No.\
SE 1836/3-2). We acknowledge the support by NewCompStar COST Action
MP1304.

\bibliographystyle{JHEP}
\bibliography{springer}{}
\end{document}